\documentclass[12pt,epsfig]{article}
\pdfoutput=1
\usepackage{graphicx}

\def\beq{\begin{equation}}
\def\eeq#1{\label{#1}\end{equation}}
\def\eeqn{\end{equation}}
\def\beqa{\begin{eqnarray}}
\def\eeqa#1{\label{#1}\end{eqnarray}}
\def\eeqan{\end{eqnarray}}
\def\CR{\nonumber \\ }
\def\leqn#1{(\ref{#1})}

\def\to{\rightarrow}

\def\stacksymbols #1#2#3#4{\def\theguybelow{#2}
    \def\vp{\lower#3pt}
    \def\sp{\baselineskip0pt\lineskip#4pt}
    \mathrel{\mathpalette\intermediary#1}}

\def\intermediary#1#2{\vp\vbox{\sp
     \everycr={}\tabskip0pt
     \halign{$\mathsurround0pt#1\hfil##\hfil$\crcr#2\crcr
              \theguybelow\crcr}}}

\def\gsim{\stacksymbols{>}{\sim}{2.5}{.2}}
\def\lsim{\stacksymbols{<}{\sim}{2.5}{.2}}

\begin{document}

\begin{titlepage}

\begin{center}
{\Huge \bf Fine-Tuning Implications of Direct}
\vskip0.5cm
{\Huge \bf Dark Matter Searches in the MSSM} 

\vskip.2cm
\end{center}
\vskip1cm

\begin{center}
{\bf Maxim Perelstein and Bibhushan Shakya} \\
\end{center}
\vskip 8pt

\begin{center}
	{\it Laboratory of Elementary Particle Physics, \\
	     Physical Sciences Building\\
	     Cornell University, Ithaca, NY 14853, USA } \\

\vspace*{0.3cm}

{\tt  mp325@cornell.edu, bs475@cornell.edu}
\end{center}

\vglue 0.3truecm

\begin{abstract}
\noindent We study theoretical implications of direct dark matter searches in the minimal supersymmetric standard model (MSSM). We assume that no accidental cancellations occur in the spin-independent elastic neutralino-quark scattering cross section, but do not impose any relations among the weak-scale MSSM parameters. We show that direct detection cross section below $10^{-44}$ cm$^2$ requires the lightest supersymmetric particle (LSP) neutralino to be close to either a pure gaugino or pure Higgsino limit, with smaller cross sections correlated with smaller admixture of the subdominant components. The current XENON100 bound rules out essentially all models in which the lightest neutralino has the Higgsino fraction between 0.2 and 0.8. Furthermore, smaller direct detection cross sections correlate with stronger fine-tuning in the electroweak symmetry breaking sector. In the gaugino LSP scenario, the current XENON100 bound already implies some fine-tuning: for example, at least 10\% tuning is required if the LSP mass is above 70 GeV. In both gaugino and Higgsino LSP scenarios, the direct dark matter searches currently being conducted and designed should lead to a discovery if no accidental cancellations or fine-tuning at a level below 1\% is present. 

\end{abstract}

\end{titlepage}

\section{Motivation and Philosophy}

The Minimal Supersymmetric Standard Model (MSSM) is a well-motivated and extensively studied extension of the Standard Model (SM) at the electroweak scale. Originally introduced as a way to solve the gauge hierarchy problem of the SM, the model turned out to have many other intriguing features. In particular, the MSSM contains an attractive particle dark matter candidate, the lightest neutralino~\cite{SUSY_DM,DMreviews}. If such a neutralino is the lightest supersymmetric particle (LSP) and R-parity is conserved, it is stable, and its thermal abundance naturally falls in the range indicated by the observed dark matter density.

Probing the microscopic nature of dark matter requires observing interactions of individual dark matter particles with ordinary matter via scattering or annihilation. An extensive experimental effort toward this goal is currently under way. In particular, direct detection techniques, which attempt to detect collisions between ambient dark matter particles and nuclei in the target, have seen significant advances in sensitivity with experiments such as CDMS~\cite{cdms}, EDELWEISS~\cite{edel}, and most recently XENON100~\cite{x100old,x100}. So far, the results of these searches are null: no evidence for dark matter has been observed.\footnote{DAMA~\cite{dama} and CoGeNT~\cite{cogent} collaborations observed effects inconsistent with known backgrounds, which may be due to dark matter scattering. If so, the MSSM dark matter is disfavored, since it falls under the minimal WIMP framework which does not provide a good simultaneous fit to these experiments and XENON100~\cite{x100}. At the moment, however, the experimental situation is quite confusing, and we will not take DAMA and CoGeNT into account in this study.} The goal of this paper is to investigate the implications of these results, and improvements that are likely to come in the next few years, for the MSSM, with the assumption that the dark matter is made entirely of neutralinos.  

In the standard minimal WIMP framework, which is applicable to the MSSM dark matter throughout the model parameter space, the results of direct detection searches are presented as upper bounds on (or perhaps, in the future, measurements of) the spin-independent neutralino-proton elastic scattering cross section at zero momentum exchange,\footnote{Converting the actually measured event rates into a cross-section bound or measurement requires a number of assumptions, such as local dark matter density (see e.g.\cite{dmdensity}) and velocity distributions (see e.g.\cite{veldist}), isospin symmetry of WIMP-nucleon couplings (see e.g.\cite{isospin}), etc. We will not discuss the potential uncertainties introduced by the conversion in this paper.} which we will call ``direct detection cross section" for short. As usual with the MSSM, the theoretical prediction for this cross section depends on a large number of model parameters. The traditional approach is to reduce the number of parameters by assuming relations among them, typically at the high energy scale ({\it e.g.} mSUGRA/cMSSM). Even then, the cross section predictions vary widely depending on the parameters, and the results are usually presented as scatter plots resulting from scanning the parameters within some broad ranges. While such plots provide a useful target for experiments, much potentially interesting information is missing. A typical scatter plot shows the cross section varying over several orders of magnitude, and it is not clear what features of the model correspond to points with higher, or lower, cross sections. Can any qualitative statements about the MSSM be made given the cross section bound of $10^{-44}$ cm$^2$ (roughly corresponding to the current XENON100 result), or some lower future bound? This will be the main focus of this paper. In particular, we will demonstrate a correlation between the direct detection cross section and the amount of fine-tuning in the electroweak sector: roughly speaking, model points with lower direct detection cross sections are more fine-tuned.

\begin{figure}[t]
\begin{center}
\centerline {
\includegraphics[width=4.5in]{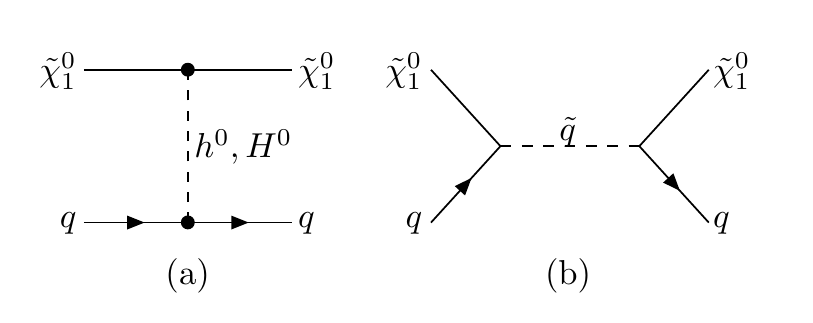}
}
\caption{Feynman diagrams contributing to spin-independent elastic scattering of the neutralino dark matter particle on a nucleon in the MSSM.}
\label{fig:Fdiagrams}
\end{center}
\end{figure}

In order to be as general as possible, we will treat all weak-scale MSSM parameters as independent, without assuming any relations among them. The tree-level processes contributing to the direct detection cross section are shown in Fig.~\ref{fig:Fdiagrams}. The key assumption underlying our analysis is that {\it no accidental cancellations take place} among various contributions to the direct detection cross section in the MSSM. By ``accidental", we mean a cancellation which is exact only on a measure-zero hypersurface inside the full MSSM parameter space. Equivalently, an accidental cancellation is indicated by an anomalous sensitivity of the cross section to the MSSM parameters (measured, for example, by its logarithmic derivative) along at least one direction in the parameter space. In particular, any cancellation between the $s$- and $t$-channel diagrams in Fig.~\ref{fig:Fdiagrams} would be accidental, since they depend on different sets of MSSM parameters.\footnote{Of course, different MSSM parameters may be related once the SUSY-breaking sector is understood, so that a cancellation that appears accidental from the weak-scale point of view may in fact be natural in the full theory. Such a situation, however, appears extremely unlikely in the particular situations where we apply the ``no accidental cancellation rule" in this study. For example, a cancellation between the $s$- and $t$-channel diagrams in Fig.~\ref{fig:Fdiagrams}  would require a complicated relation involving squark and gaugino soft masses, the $\mu$ parameter, $\tan\beta$, and the Higgs mass terms. It is very difficult to imagine a SUSY-breaking model producing such a relation.} Thus, for making qualitative statements, it is sufficient to consider only one of the diagram classes; the other one will, at worst, produce an order-one correction to the cross section. We will focus on the $t$-channel diagrams, Fig.~\ref{fig:Fdiagrams} (a). We make this choice because three of the five MSSM parameters which enter these diagrams, $\mu$, $\tan\beta$, and $m_A$, also enter the tree-level prediction for the $Z$ mass. In this way, the direct detection cross section is connected to electroweak symmetry breaking.

When comparing with experimental data, we will assume that the local dark matter density is at its canonical value, $0.3$ GeV/cm$^3$, and that dark matter is made entirely of the MSSM neutralinos. However, the main results of the paper concern theoretical predictions for direct detection cross section and are independent of this assumption. Moreover, in most of the analysis, we will {\it not} impose the constraint that the neutralino relic density predicted by the standard thermal decoupling calculation be consistent with observations. This is motivated partly by the desire to keep the analysis as general as possible: Statements made without the relic density constraint would be applicable even if there is non-thermal production of neutralinos, late inflation, other deviations from standard FRW cosmological evolution, {\it etc}. More technically, it is motivated by the fact that the object we focus on, the $t$-channel direct detection cross section, depends on just five MSSM parameters, while the relic density is a function of many more. Imposing this constraint would thus greatly complicate the analysis.
(As an exception, we will impose a mild version of the relic density constraint in the part of the analysis dealing with Higgsino dark matter, where this will be required to make an interesting statement about fine-tuning.)    

Before we proceed, let us remark on a few related analyses in the literature. The connection between direct detection cross section and fine-tuning in the electroweak sector was noted, in the mSUGRA context, in Ref.~\cite{GPMM} (see also~\cite{sa,DG}). A study of this connection in a general MSSM framework similar to ours, focusing on the bino-LSP scenario, appeared in Ref.~\cite{KN}. A recent study of ``generic" direct detection cross sections in a general MSSM framework was presented in Ref.~\cite{FS}. While our approach is similar, our setup is even more general: in particular, we do not require gaugino mass unification, and (for the most part) do not impose the thermal relic density constraint. Electroweak fine-tuning, which did not enter the analysis of~\cite{FS}, plays the central role in our discussion. We also discuss the impact of the recent XENON100 bound in the general MSSM framework, with the main assumption being the absence of accidental cancellations. This discussion complements Refs.~\cite{akula,strumia,xenon_msugra} which interpreted the XENON100 result within mSUGRA and other constrained MSSM frameworks.  
 
The rest of the paper is organized as follows. In section~\ref{sec:setup}, we set the notation and collect the main formulas used in the analysis. The main results of the paper are contained in section~\ref{sec:partial}, 
where we present a set of scatter plots demonstrating correlations between the direct detection cross section and physical quantities such as Higgsino fraction of the neutralino and the amount of fine-tuning in the electroweak symmetry breaking. These plots assume real, positive MSSM parameters. This assumption 
is lifted in section~\ref{sec:full}, where negative and complex soft masses are considered, and it is shown that the interesting correlations persist once the points with accidental cancellations in the cross section are eliminated. Finally, we conclude in section~\ref{sec:conc}. 
 
\section{Analysis Setup}
\label{sec:setup}

The direct detection cross section has the form~\cite{DMreviews}
\beq
\sigma = \frac{4m_r^2 f_p^2}{\pi}\,,
\eeq{xsec}
where $m_r$ is the neutralino-proton reduced mass and
\beq
\frac{f_p}{m_p}\,=\,\sum_{q=u,d,s} f_{T_q}^{(p)} A_q \,+\, \frac{2}{27}\, f_{TG}^{(p)} \sum_{q=c,b,t} A_q.
\eeq{fp} 
The nuclear formfactors are defined by
\beq
f_{T_q}^{(p)} = \frac{\langle m_q \bar{q}q \rangle}{m_p}\,,
\eeq{fnucs}
and
\beq
f_{TG}^{(p)} = 1 - \sum_{q=u,d,s} f_{T_q}^{(p)}\,.
\eeq{fglue}
In our numerical work, we will use the following values~\cite{nucs}:
\beq
f_{Tu}^{(p)} = 0.08,~~f_{Td}^{(p)} = 0.037,~~f_{Ts}^{(p)} = 0.34.
\eeq{fnum}
It is well known that there is a significant uncertainty on these formfactors, in particular $f_s$~\cite{nucs}. The uncertainty could easily be incorporated in our study; however, since our main interest is in qualitative trends rather than precise quantitative statements, we will not do so here. The dependence of the cross section on the underlying particle physics model is contained in the coefficients $A_q$, which in the MSSM at tree level are computed with the Feynman diagrams in Fig.~\ref{fig:Fdiagrams}. As explained above, we will focus on the $t$-channel contribution, given by~\cite{As}
\beqa
A_i &=& -\frac{g}{4m_WB_i}\Bigl[ \left( \frac{D_i^2}{m_h^2}+\frac{C_i^2}{m_H^2} \right) \, {\rm Re}
\left[ \delta_{2i}(gZ_{\chi2}-g^\prime Z_{\chi1})\right]  \CR & & + C_i D_i \left( \frac{1}{m_h^2}-\frac{1}{m_H^2} \right)\, {\rm Re} \left[ \delta_{1i}(gZ_{\chi2}-g^\prime Z_{\chi1})\right] \Bigr] \,,
\eeqa{A_i}
where for up-type quarks
\beq
B_u = \sin\beta\,,~~C_u = \sin\alpha\,,~~D_u = \cos\alpha\,,~~\delta_{1u}=Z_{\chi3}\,,~~\delta_{2u}=Z_{\chi4}\,;
\eeq{defs_up}
while for down-type quarks
\beq
B_d = \cos\beta\,,~~C_d = \cos\alpha\,,~~D_d = -\sin\alpha\,,~~\delta_{1d}=Z_{\chi4}\,,~~\delta_{2d}=-Z_{\chi3}\,.
\eeq{defs_d}
The notation here is consistent with that used, for example, in Martin's review~\cite{Martin}; in particular, the lightest neutralino is
\beq
\tilde{\chi}^0_1 = Z_{\chi 1} \tilde{B} + Z_{\chi 2} \tilde{W}^3 + Z_{\chi 3} \tilde{H}^0_d + Z_{\chi 4} \tilde{H}^0_u\,,
\eeq{chi01}
and $h$ and $H$ are the two CP-even Higgs bosons ($m_h<m_H$). The coefficients $A_i$ can be expressed in terms of only five MSSM parameters:
\beq
p_i = (M_1, M_2, \mu, \tan\beta, m_A)\,, 
\eeq{5pars}
where $M_1$, $M_2$, and $\mu$ appear in Eq.~\leqn{A_i} implicitly via the neutralino mixing matrix elements. These parameters will always be defined at the weak scale; since no unification or any other relation among the parameters is assumed, we do not need to consider their renormalization group evolution. In general, the parameters $p_i$ are complex; however it can be shown (see, for example, Ref.~\cite{welltemp}) that only two phases are physical:
\beq
\varphi_1=\arg( \mu M_1\sin2\beta),~~~\varphi_2=\arg(\mu M_2\sin2\beta )\,.
\eeq{phys_phases}
These phases are constrained by measurements of electric dipole moments (EDMs)~\cite{EDMs}, although maximal phases are allowed if squarks and sleptons are very heavy~\cite{bigphases}.
The light Higgs mass $m_h$ can be expressed in terms of the parameters in~\leqn{5pars} at tree level. It is of course well known that a large loop correction is required to satisfy the 
LEP-2 lower bound on $m_h$; this correction is dominated by the top and stop loops and including it would bring in a few additional MSSM parameters into the game. In this study, we avoid doing this by simply fixing $m_h$ at a fixed value consistent with LEP-2, $m_h=120$ GeV. (In the MSSM, the upper bound on $m_h$ is about 135 GeV; a variation of $m_h$ in the allowed range does not have a strong effect on the direct detection cross section.) In other words, we assume that for {\it any} set of $p_i$'s, the other MSSM parameters can be chosen so that $m_h=120$ GeV. 

The other quantity of interest for us is the amount of fine-tuning in the electroweak symmetry breaking (EWSB) sector of the model. A tree-level analysis will suffice. (The discussion below is taken from Ref.~\cite{golden}.) The key formula is the relation of the $Z$ boson mass to the MSSM Lagrangian parameters: 
\beq
m_Z^2 \,=\, - m_u^2 \left( 1-\frac{1}{\cos 2 \beta} \right) - 
m_d^2 \left( 1+\frac{1}{\cos 2 \beta} \right)-2|\mu|^2 \,,
\eeq{zmass}
where $m_u^2$ and $m_d^2$ are the Lagrangian masses for the up-type and down-type Higgs doublets, and
\beq
\sin 2 \beta  = \frac{2 b}{m_u^2 + m_d^2+2|\mu|^2}\,.
\eeq{sin} 
We quantify fine-tuning by computing
\beq
\delta(\xi)\,=\,\left| \frac{\partial\log m_Z^2}{\partial\log \xi}\right|,
\eeq{ftpars}
where $\xi=m_u^2, m_d^2, b, \mu$ are the relevant Lagrangian parameters. (This is analogous to the fine-tuning measure introduced by Barbieri and Guidice~\cite{BG}, although here it is applied to weak-scale, rather than Planck/GUT-scale, MSSM parameters.)
Using the well-known tree-level relations to express $m_u^2$ and $m_d^2$ in terms of the parameters listed in~\leqn{5pars}, we obtain~\cite{golden}
\beqa
\delta(\mu) &=& \frac{4\mu^2}{m_Z^2}\,\left(1+\frac{m_A^2+m_Z^2}{m_A^2}
\tan^2 2\beta \right), \CR 
\delta(b) &=& \left( 1+\frac{m_A^2}{m_Z^2}\right)\tan^2 2\beta, \CR
\delta(m_u^2) &=& \left| \frac{1}{2}\cos2\beta +\frac{m_A^2}{m_Z^2}\cos^2\beta
-\frac{\mu^2}{m_Z^2}\right|\times\left(1-\frac{1}{\cos2\beta}+
\frac{m_A^2+m_Z^2}{m_A^2} \tan^2 2\beta \right), \CR 
\delta(m_d^2) &=& \left| -\frac{1}{2}\cos2\beta +\frac{m_A^2}{m_Z^2}\sin^2\beta
-\frac{\mu^2}{m_Z^2}\right|\times\left|1+\frac{1}{\cos2\beta}+
\frac{m_A^2+m_Z^2}{m_A^2} \tan^2 2\beta \right|, \CR 
\eeqa{ders}
where we assumed $\tan\beta>1$. The overall fine-tuning $\Delta$ is defined
by adding the four $\delta$'s in quadruture; values of $\Delta$ far above 1
indicate fine-tuning. The qualitative behavior of fine-tuning is easy to understand by taking the limit of large 
$\tan\beta$, where the parameters $\delta(m_u^2)$
and $\delta(m_d^2)$ are small, and 
\beq
\delta(\mu) \approx \frac{4\mu^2}{m_Z^2},~~~\delta(b) \approx \frac{4m_A^2}{m_Z^2 \tan\beta}\,.
\eeq{conlarget}
Thus, increasing $\mu$, and to a lesser extent $m_A$, requires fine-tuning. On the other hand, as $\beta$ approaches 
$\pi/4$, the factors of $1/\cos2\beta$ and $\tan2\beta$, present in all four 
$\delta$'s, become large, and as a result the model is always fine-tuned 
for $\tan\beta\lsim 2$. 

Before proceeding, let us clarify the following point.\footnote{We are grateful to D.~Ghilencea for a question that prompted this clarification.} The definition of EWSB fine-tuning used here captures the fine-tuning between the various parameters that enter the tree-level relation, Eq.~\leqn{zmass}. It does {\it not} include the fine-tuning between tree-level and loop-level contributions to the $Z$ mass. In particular, it is well known that, in the MSSM, the corrections to $m_u^2$ from top loops 
are large in the parameter region where the Higgs is heavy enough to satisfy the LEP-2 constraint~\cite{LH}. This necessitates a fine-tuning in the $Z$ mass which is, at best, of order a few \%. This fine-tuning has a very different physical origin from the tuning required to satisfy dark matter direct detection bounds; in particular, the parameters that predominantly determine the loop contribution to $m_u^2$ (third-generation squark soft masses and $A$-terms) play no role in the direct detection cross sections. The numerical values of fine-tuning we calculate refer {\it only} to the tree-level tuning required by dark matter bounds alone. Roughly speaking, one can estimate the total fine-tuning in the EWSB sector by adding these two contributions in quadrutures; we will not do so explicitly in this paper.

Our strategy for the rest of the paper is as follows: We will perform scans over the five parameters $p_i$, computing the direct detection cross section and the fine-tuning measure discussed above for each point, and study the correlation between these two quantities as well as with other relevant parameters such as the LSP mass, Higgsino/gaugino fractions, etc. We will then explain physical reasons for each observed interesting correlation. 

\section{Results: Real, Positive Parameters}
\label{sec:partial}

We begin by performing a restricted scan in which we assume that all five $p_i$ parameters are {\it real and positive}. All the correlations that we found show up most clearly in this scan, making it a good place to start the discussion. In more complete scans, the correlations persist, but are somewhat obscured by the possibility of accidental cancellations within the $t$-channel contribution to the direct detection cross section. This will be discussed in detail in the next section. 

A set of 10$^5$ MSSM points was generated, distributed randomly, uniformly in $\log M_1$, $\log M_2$, $\log\mu$, $\log m_A$, and $\tan\beta$, within the following scan boundaries:
\beqa
& &M_1 \in [10, 10^4]~{\rm GeV};~~~~~M_2 \in [80, 10^4]~{\rm GeV}; \CR
& &\mu \in [80, 10^4]~{\rm GeV};~~~~~m_A \in [100, 10^4]~{\rm GeV}; \CR
& & \tan\beta \in [2,50]\,.
\eeqa{scanbound_1}
We compute the neutralino and chargino masses for each point in the scan, and exclude points with at least one chargino below 100 GeV (excluded by LEP-2), as well as those where the lightest chargino mass is below the lightest neutralino mass. We do not impose any other experimental constraints, since they depend on the MSSM parameters beyond the five $p_i$ being scanned here, and thus generically can be satisfied by varying those parameters for any given $p_i$. The scatter plots in this section are based on the 73064 points that pass these constraints.

\subsection{Higgsino Fraction Constraint}
\label{sec:Hfrac}

\begin{figure}[t]
\begin{center}
\centerline {
\includegraphics[width=3.3in]{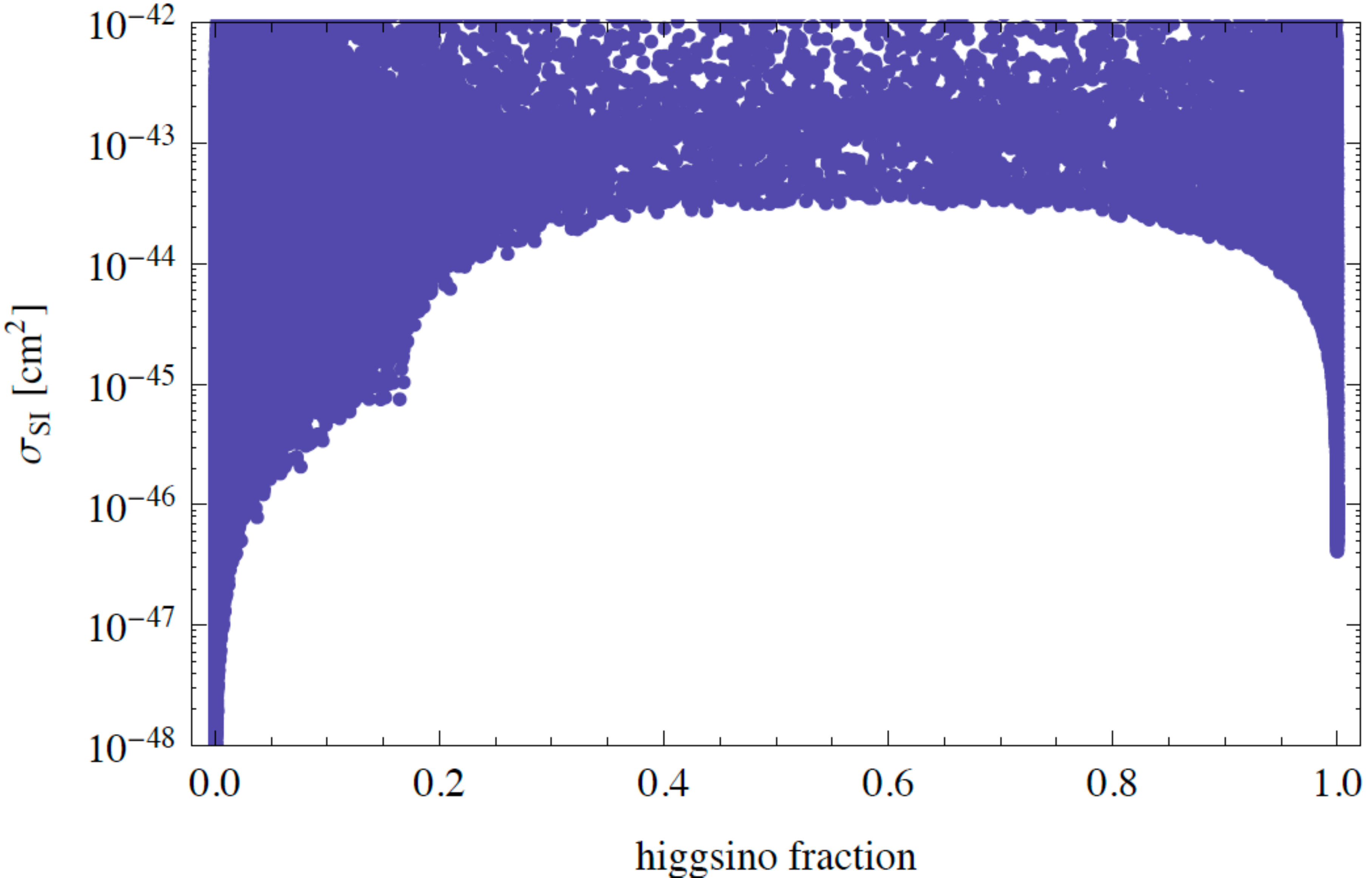}
\includegraphics[width=3.3in]{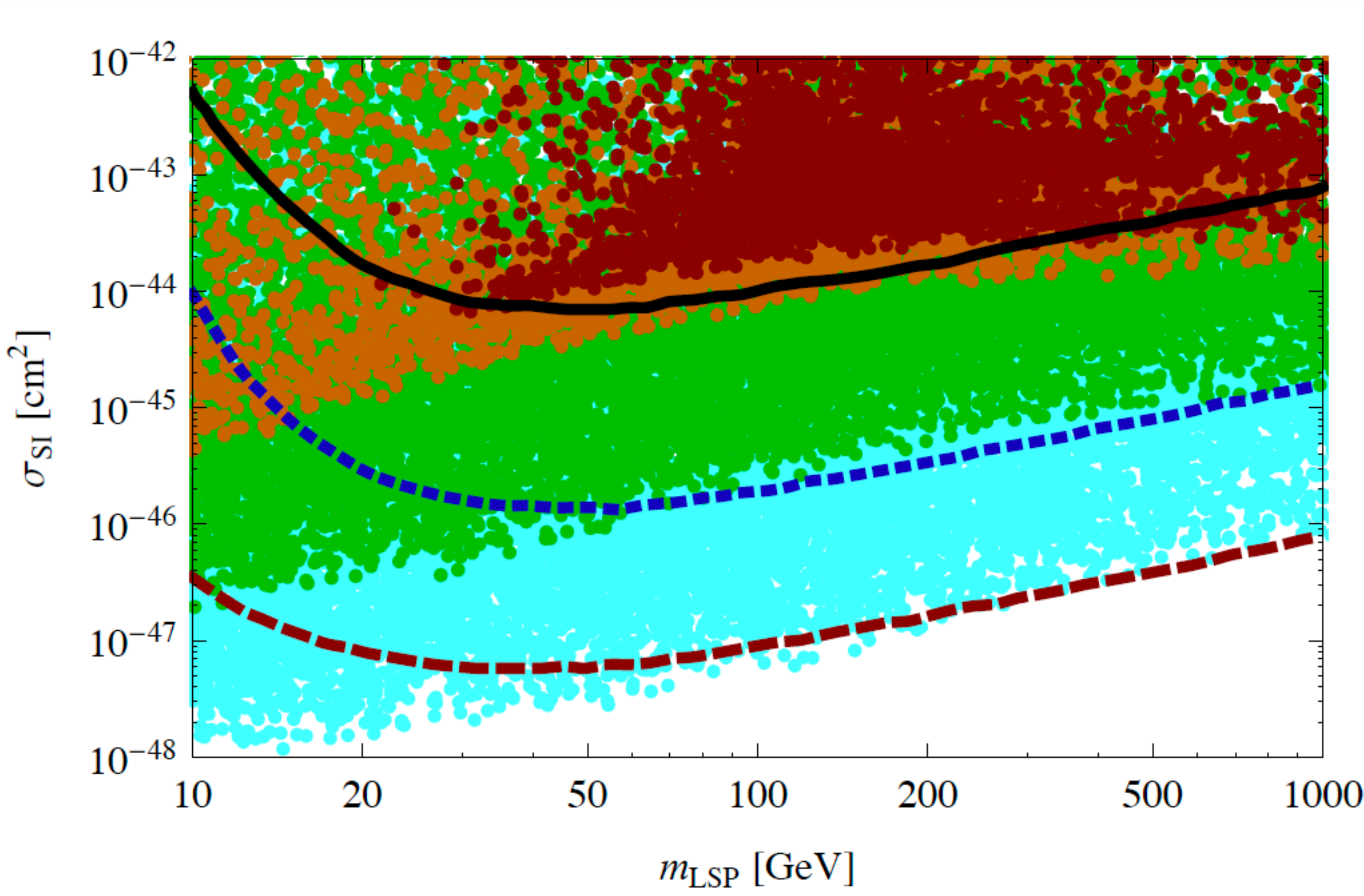}
}
\caption{Left panel: Direct detection cross section vs. Higgsino fraction of the neutralino.
Right panel: Direct detection cross section vs. the dark matter particle mass, for points with purity above 0.2 (red), between 0.1 and 0.2 (orange), 0.01 and 0.1 (green), and $10^{-3}$ and 0.01 (cyan). The lines correspond to the XENON100 100 days exclusion limit~\cite{x100} (black/solid), and the projected sensitivities of the XENON100 upgrade~\cite{x100F,xtalk} (blue/dotted) and XENON-1T~\cite{xtalk} (red/dashed).  Real, positive values of the scanned MSSM parameters are assumed.}
\label{fig:hfraction}
\end{center}
\end{figure}

Our first result concerns the correlation between the direct detection cross section and the Higgsino fraction of the neutralino, defined as
\beq
F_H = |Z_{\chi3}|^2 + |Z_{\chi4}|^2\,. 
\eeq{Hfraction}
As is clear in Fig.~\ref{fig:hfraction}, a direct detection cross section limit below (a few)$\times 10^{-44}$ cm$^2$ puts a constraint on the Higgsino fraction, requiring that it be close to either 0 (the ``pure gaugino" case) or 1 (the ``pure Higgsino" case). It is convenient to define ``neutralino purity" $p$ as
\beq
p = \min (F_H, 1-F_H).
\eeq{purity}
The bound placed by the XENON100 experiment~\cite{x100} already rules out essentially all MSSM dark matter models with $p>0.2$, and most models with $p>0.1$, especially for the LSP mass above 50 GeV. \footnote{A well-known example of such a model is the ``well-tempered neutralino" scenario~\cite{welltemp}. The fact that this scenario is disfavored by XENON100 has already been noted in Ref.~\cite{strumia}.}  The proposed XENON100 upgrade~\cite{x100F,xtalk} will be able to probe values of $p$ down to 0.01 for the LSP masses above 50 GeV, while a 1-ton upgrade~\cite{xtalk} will have a reach down to $p\approx 10^{-3}$ through most of the mass range.\footnote{Many next-generation direct dark matter searches have been proposed, such as XMASS~\cite{xmass}, LUX~\cite{lux}, and superCDMS~\cite{supercdms}. Needless to say, we use projections from the XENON collaboration simply as a benchmark, and do not mean to endorse or express a preference for a particular technology or experimental proposal. Projected sensitivities of any proposed experiment can be easily superimposed on our plots.}  If the dark matter is not discovered at that stage, the only possibility in the MSSM would be a pure gaugino or Higgsino with $<0.1$\% admixture of the other components. 

The physical origin of this constraint is easy to understand. In the gauge eigenbasis for neutralinos, the 
neutralino-neutralino-Higgs couplings have the form $(g \tilde{W^3} +g^\prime \tilde{B}) \tilde{H} H$; there are no gaugino-gaugino-Higgs or Higgsino-Higgsino-Higgs couplings in the MSSM. In the mass basis, the couplings have the form 
\beqa
\tilde{\chi}^0\tilde{\chi}^0h:& &~(gZ_{\chi2}-g^\prime Z_{\chi1})(\cos\alpha Z_{\chi4}+\sin\alpha Z_{\chi3} )\,,\CR
\tilde{\chi}^0\tilde{\chi}^0H:& &~(gZ_{\chi2}-g^\prime Z_{\chi1})(\sin\alpha Z_{\chi4}-\cos\alpha Z_{\chi3} )\,.
\eeqa{nnh}
If the $\tilde{\chi}^0\tilde{\chi}^0h$ is of its natural size ({\it i.e.} no accidental cancellations or small mixing angles are present), the direct detection cross section from $t$-channel Higgs exchange is of order (a few)$\times 10^{-44}$ cm$^2$ or above. Barring accidental cancellations, the only way to obtain a smaller cross section is to suppress this coupling by choosing the LSP to be an almost pure gaugino or Higgsino, which is precisely what is seen in Fig.~\ref{fig:hfraction}.

\subsection{Gaugino Dark Matter and Electroweak Fine-Tuning}
\label{sec:EWSB_FT}

To discuss the connection between direct detection cross section and EWSB fine-tuning, it is useful to divide the scan points into two sets: the points where $\mu<M_1$ and $\mu<M_2$, and the rest. We will refer to the first set of points as the ``Higgsino LSP" sample, while the second set will be called ``gaugino LSP" sample. Of course, these names are not precise, since each sample contains points with $\mu\sim M_{1,2}$ where the LSP is a roughly equal mixture of the two; however, as we saw above, such points always have high direct detection cross sections and will not influence the fine-tuning discussion. (Most of these points are in any case already ruled out by XENON100, though for simplicity we will not impose this constraint here.)  

\begin{figure}[tb]
\begin{center}
\centerline {
\includegraphics[width=4in]{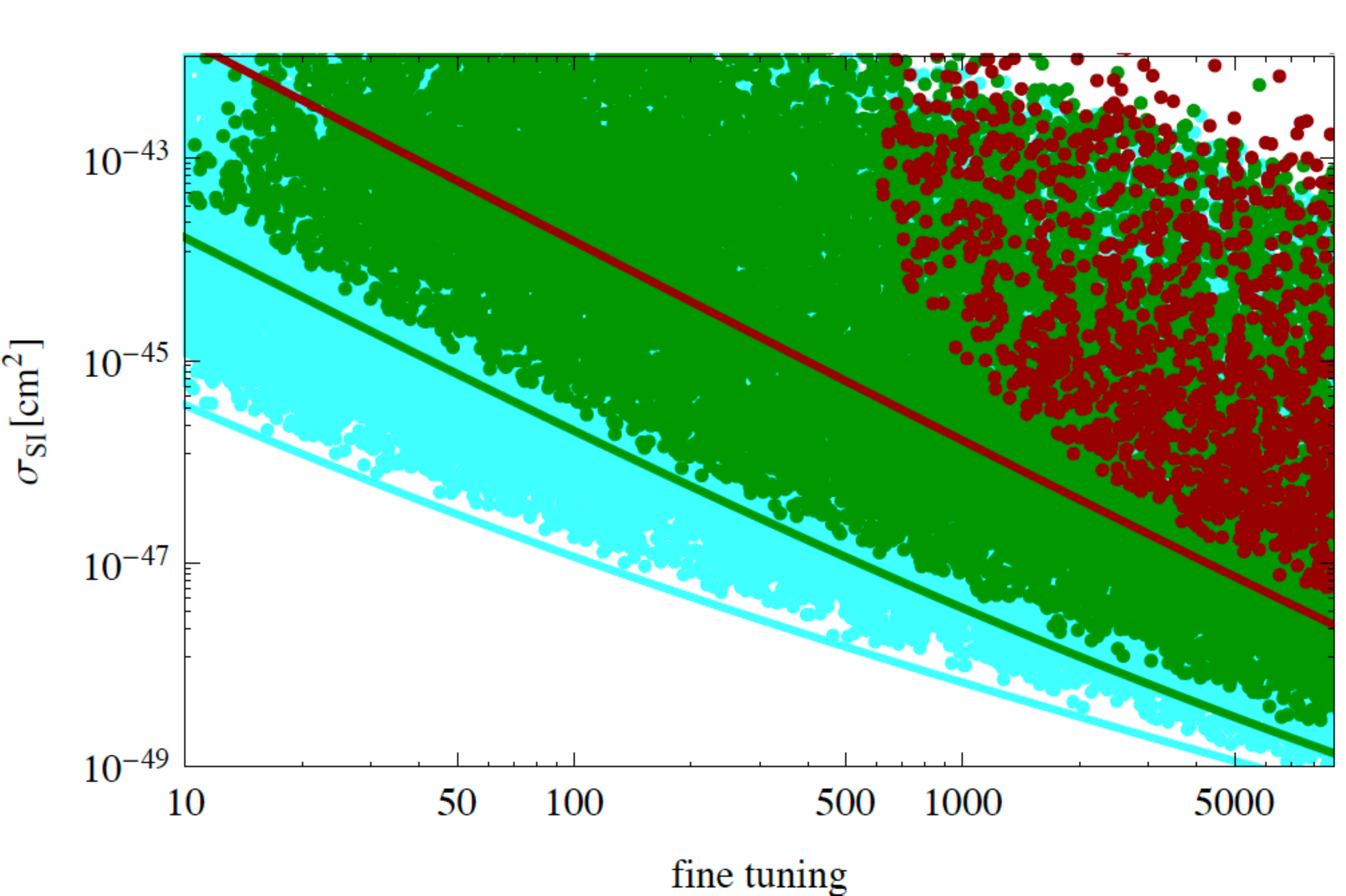}
}
\caption{Direct detection cross section vs. fine-tuning of electroweak symmetry breaking, for gaugino-like neutralino (real, positive values of $p_i$, $i=1\ldots 5$, are assumed). Cyan, green and red points correspond to the dark matter particle mass in the intervals $[10, 100)$, $[100, 1000)$, and $[1000, 10^4]$ GeV, respectively. The cyan, green and red lines show the analytic lower bound in Eq.~\leqn{FTanal}, with $M_{\rm LSP}=10, 100, 1000$ GeV, respectively. Real, positive values of the MSSM parameters are assumed.}
\label{fig:EWtuning}
\end{center}
\end{figure}

Let us first consider the gaugino LSP sample. The correlation between the direct detection cross section and the amount of fine-tuning in the EWSB sector (quantified by $\Delta$ defined in Sec.~\ref{sec:setup}) in this sample is shown in Fig.~\ref{fig:EWtuning}. For a given amount of fine-tuning, the direct detection cross section cannot be reduced below a certain lower bound, with smaller cross sections possible only for more finely-tuned models. The physical origin of this correlation is again simple: as we saw above, the cross section can only be reduced by reducing the Higgsino admixture in the LSP, and the only way to achieve this is to raise $\mu$. But doing so increases the fine-tuning, as is easily seen from Eq.~\leqn{zmass} or Eq.~\leqn{conlarget}. It should also be mentioned that the sparsely populated region in the top right corner is merely a consequence of our choice of scan boundaries and distribution (points in this region require both large $\mu$ to be greatly fine-tuned and comparably large $M_1$ or $M_2$ to have large cross sections, which is extremely unlikely given our scan points are log distributed in these parameters).  

Also plotted in Fig.~\ref{fig:EWtuning} is an analytic expression for the {\it minimal} direct detection cross cross section possible for a given amount of fine-tuning. For a given $\tan\beta$ and LSP mass, it is given by (for derivation, see Appendix~\ref{app:FTanal}): 
\beq
\sigma_{\rm min} \,=\, (1.2\times 10^{-42}~{\rm cm}^2) \,\left( \frac{120~{\rm GeV}}{m_h}\right)^4 \frac{1}{\Delta}\,\left( \frac{1}{\tan\beta} \,+\, \frac{1}{\sqrt{\Delta}}\frac{M_{\rm LSP}}{m_Z}\right)^2\,.
\eeq{FTanal}
It is clear that the lowest possible cross section for fixed $\Delta$ requires the highest possible $\tan\beta$ and the lowest possible $M_{\rm LSP}$. In the plot in Fig.~\ref{fig:EWtuning}, we used $\tan\beta=50$, corresponding to the upper boundary of the scan. Another noteworthy feature is that the lowest possible direct detection cross sections occur in the Higgs decoupling limit, $m_A\gg m_Z$.  

\begin{figure}[tb]
\begin{center}
\centerline {
\includegraphics[width=3.3in]{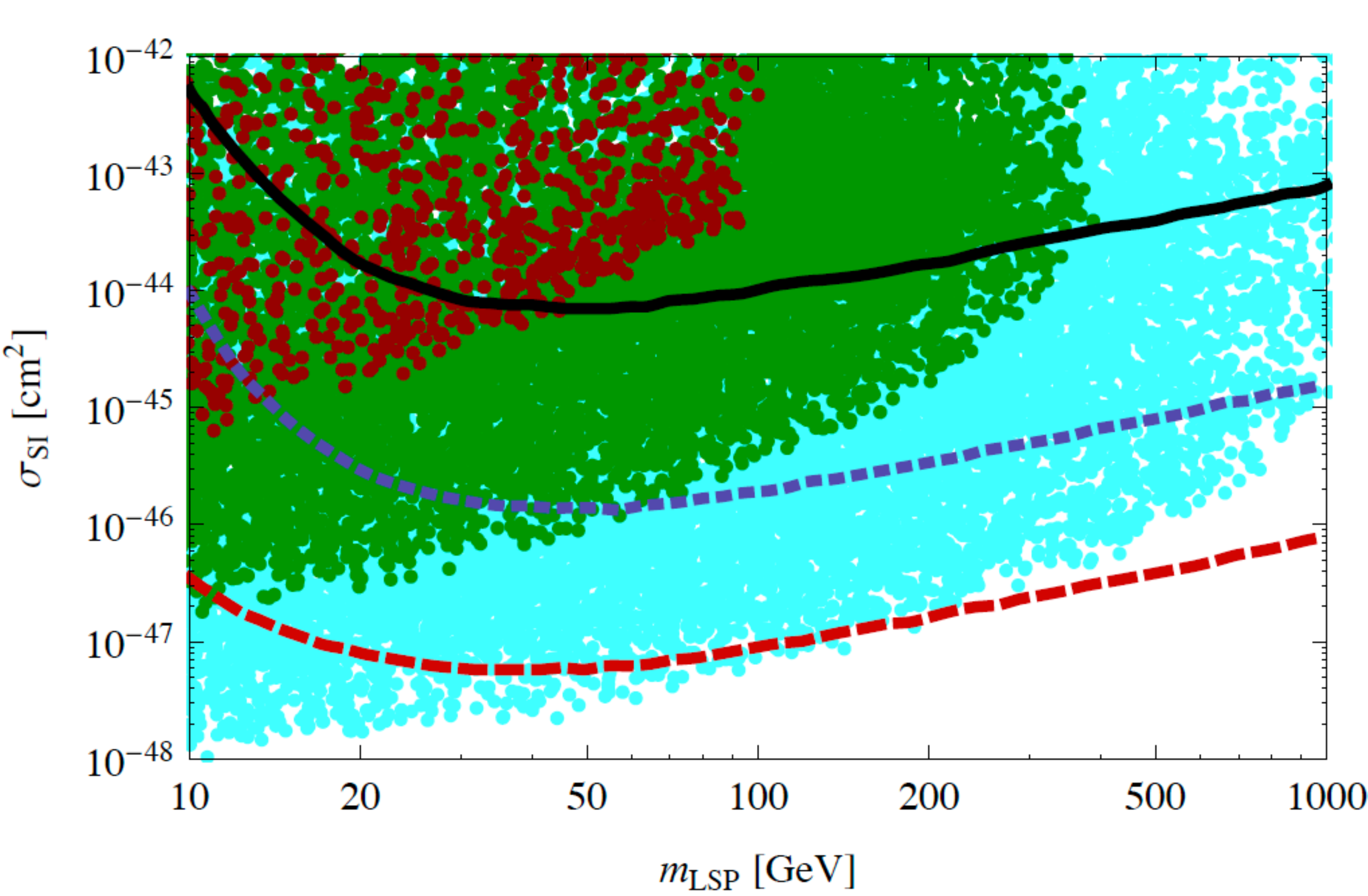}
\includegraphics[width=3.3in]{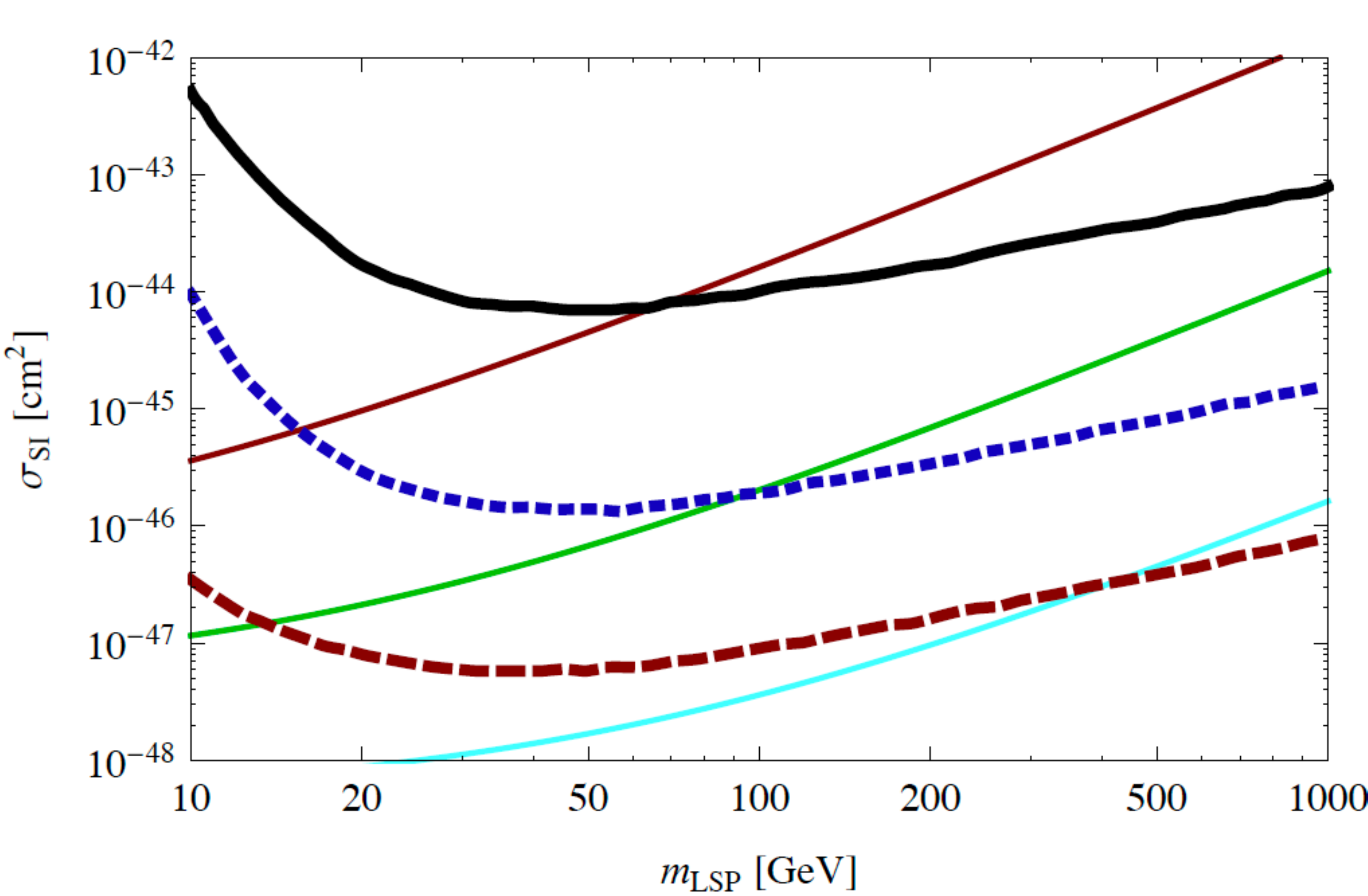}
}
\caption{Left panel: Direct detection cross section vs. dark matter particle mass, for gaugino-like neutralino.
Red, green and cyan points correspond to EWSB fine-tuning in the intervals $(0, 10)$; $[10, 100)$; $[100,1000)$, respectively. Right panel: Minimal direct detection cross section compatible with EWSB fine-tuning of 10 (red line), 100 (green line), and 1000 (cyan line) according to Eq.~\leqn{FTanal}. Current and projected XENON bounds are superimposed (notation as in Fig.~\ref{fig:hfraction}). Real, positive values of the MSSM parameters are assumed.}
\label{fig:MLSP_tuning}
\end{center}
\end{figure}

The practical implication of the correlation in Fig.~\ref{fig:EWtuning} is clear: in the gaugino LSP scenario, {\it a sufficiently strong bound on direct detection cross section implies a non-trivial fine-tuning in the EWSB sector of the MSSM.} The precise level where the cross section bounds become relevant for fine-tuning depends on the LSP mass. This is illustrated in Fig.~\ref{fig:MLSP_tuning}. The current XENON100 bound is already relevant: for example, fine-tuning of 1 part in 10 or less is only possible for LSP masses below about 70 GeV. Null result of the proposed XENON100 upgrade~\cite{x100F,xtalk} would imply at least 1\% fine-tuning for WIMPs at 100 GeV or above. This level of fine-tuning is similar to what is required to accommodate the non-observation of the Higgs boson at LEP-2, the famous ``little hierarchy problem"~\cite{LH} of the MSSM. A XENON-1T upgrade~\cite{xtalk} would be able to probe all models fine-tuned at 1\% level or better, and most models fine-tuned at $0.1$\% or better. No observation at the level of sensitivity where irreducible neutrino backgrounds would render further progress impossible, about 10$^{-48}$ cm$^2$~\cite{nus}, would imply fine-tuning of at least 0.1\% across the entire LSP mass range.  We stress that these statements are remarkably general: we only assumed that a mostly-gaugino LSP contributes all of the present dark matter, and that no accidental cancellations occur in the direct detection cross section. No assumptions whatsoever have been made about the SUSY-breaking model, and the neutralino thermal relic density constraint has not been imposed, so that the result is independent of cosmological history. (The discussion here also assumes that $p_i$'s are real and positive, but this restriction does not substantially affect the results, as shown in the next section.)

\subsection{Higgsino Dark Matter}
\label{sec:higgsino}

Now, let us switch our attention to the Higgsino LSP sample. It is clear that, if no additional assumptions are made, no correlation between direct detection cross section and fine-tuning exists in this case. Indeed, if $\mu\sim 100$ GeV, no fine-tuning is necessary in the EWSB, while the direct detection cross section can be suppressed by choosing $M_{1,2}\gg \mu$. Large values of $M_{1,2}$ only affect fine-tuning at the one-loop level, and these parameters can be at the multi-TeV scale without significant tuning. In this situation, the direct detection cross sections are many orders of magnitude below the current sensitivity, and can even be below the $10^{-48}$ cm$^2$ level where the neutrino background would render direct detection impossible. 

To make an interesting statement in this situation, we need to make an additional assumption. An obvious one is to demand that the LSP has the correct relic density, assuming conventional FRW cosmology and no non-thermal production. The relic density is typically determined by the LSP annihilation cross section. 
In the limit of pure Higgsino LSP, the next-to-lightest neutralino and the lightest chargino are quasi-degenerate with the LSP, and co-annihilations among these states need to be taken into account in the relic density calculation~\cite{higgsinodm}. The cross sections are typically dominated by annihilations into electroweak gauge bosons, $W$'s and $Z$'s. At tree level, these cross sections are completely determined by the same five MSSM parameters, Eq.~\leqn{5pars}, that entered our analysis of direct detection. There are, of course, other contributions to the annihilation cross section, such as the diagrams with top final states. Including these processes would introduce more parameters and complexity into our analysis. To avoid this, while still keeping open the possibility that they contribute significantly to the cross section, we impose a {\it one-sided} relic density constraint: we demand that the cross section of (co)annihilation into  $W/Z$ final states be {\it no larger} than what is required to obtain the observed relic density, $\Omega_{\rm dm}h^2=0.110\pm 0.006$~\cite{PDG}. 

\begin{figure}[t]
\begin{center}
\centerline {
\includegraphics[width=4in]{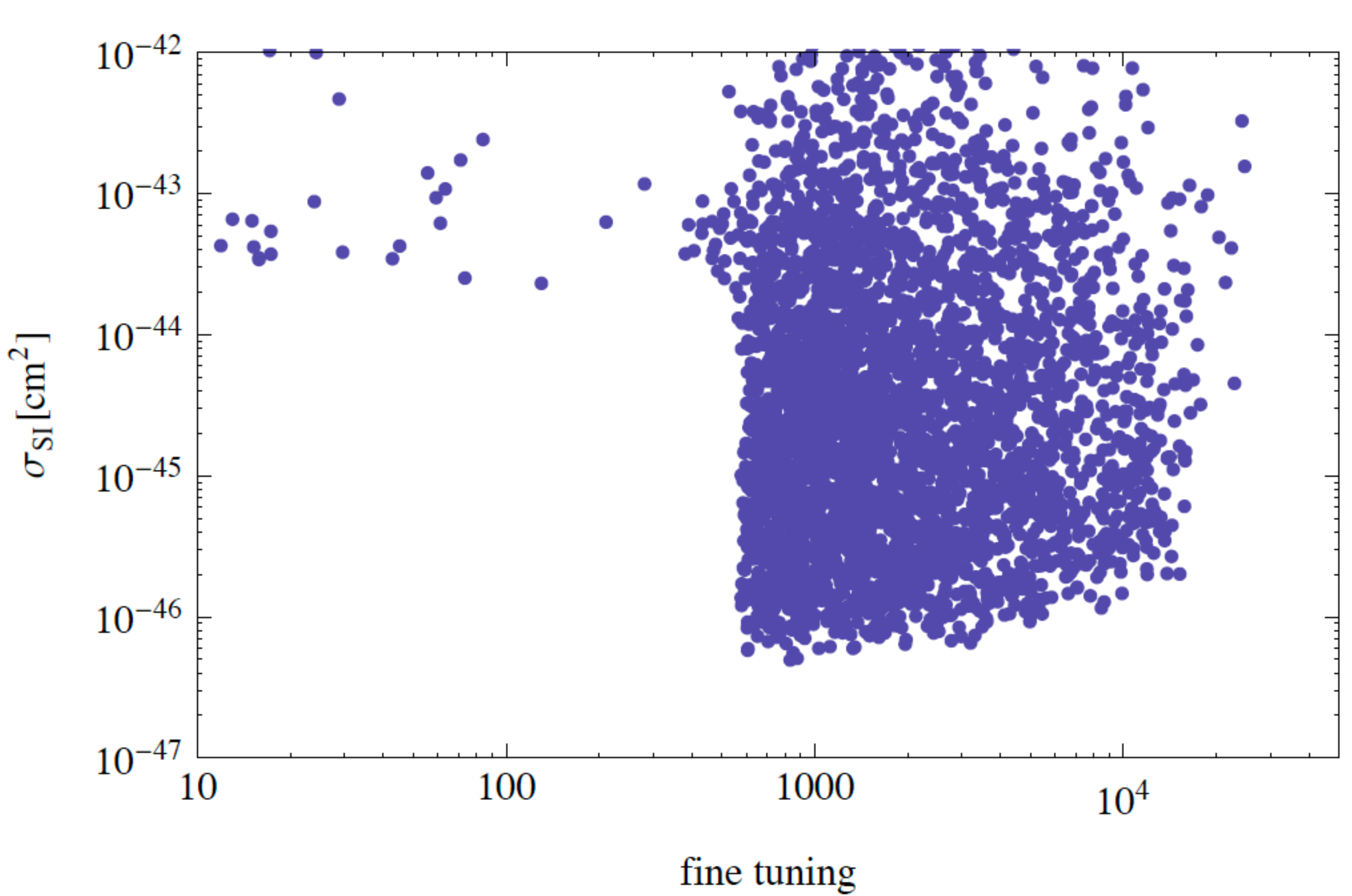}
}
\caption{Direct detection cross section vs. fine-tuning of electroweak symmetry breaking, for higgsino-like neutralino. Real, positive values of the MSSM parameters are assumed, and a one-sided thermal relic density constraint, $\Omega_{\rm pred}\geq \Omega_{\rm obs}$, is imposed, as explained in the text.}
\label{fig:higgsino}
\end{center}
\end{figure}

For each point in the Higgsino LSP sample, we performed the relic density calculation using the numerical package DarkSUSY~\cite{darksusy}. In DarkSUSY runs, all mass scales other than those in the five input parameters in~\leqn{5pars} were set to $10$ TeV, effectively eliminating annihilations into quarks and leptons. We require that the relic density calculated with DarkSUSY not be lower than 2$\sigma$ below the observed relic density. Once this constraint is imposed, a simple relation between fine-tuning and direct detection cross section emerges, seen clearly in Fig.~\ref{fig:higgsino}. If direct detection cross section is constrained to be below about $2\times 10^{-44}$ cm$^2$, the LSP must be a pure Higgsino, as we saw in section~\ref{sec:Hfrac}. For pure Higgsino, the annihilation cross section is too large, unless $\mu\gsim 1$ TeV. But such large values of $\mu$ require fine-tuing in the EWSB of at least about 0.25\%, as is easily seen from Eq.~\leqn{conlarget}. Thus, a direct detection bound of about 10$^{-44}$ cm$^2$ for a 1 TeV LSP mass would imply a ``little hierarchy problem" for Higgsino LSP. The current XENON100 bound at 1 TeV mass is about $8\times 10^{-44}$ cm$^2$, so no such statement can yet be made. The proposed XENON100 upgrade~\cite{x100F,xtalk} should reach the required sensitivity. If no signal is discovered, the Higgsino LSP scenario would be inconsistent with naturalness of EWSB at a sub-percent level.

\section{Results: Full Scan}
\label{sec:full}

In this section, we remove the requirement of real, positive $p_i$. Since complex phases of the MSSM parameters are generally constrained by measurements of EDMs, we first remove the condition $p_i>0$, but keep them real. Only two phases are physical, see Eq.~\leqn{phys_phases}, and we choose the basis where $\mu$ and the Higgs vevs are positive but $M_1$ and $M_2$ can have either sign. We generate a set of 10$^5$ MSSM points distributed randomly, uniformly in $\log |M_1|$, $\log |M_2|$, $\log\mu$, $\log m_A$, and $\tan\beta$, within the following scan boundaries:
\beqa
& &|M_1| \in [10, 10^4]~{\rm GeV};~~~~~|M_2| \in [80, 10^4]~{\rm GeV}; \CR
& &\mu \in [80, 10^4]~{\rm GeV};~~~~~m_A \in [100, 10^4]~{\rm GeV}; \CR
& & \tan\beta \in [2,50]\,.
\eeqa{scanbound_2}
The signs of $M_1$ and $M_2$ are chosen between $+$ and $-$ with equal probability. After imposing the same requirements as in the positive-only scan of section~\ref{sec:partial}, we are left with 76546 points, which are included in the scatter plots below.

\begin{figure}[h]
\begin{center}
\centerline {
\includegraphics[width=4in]{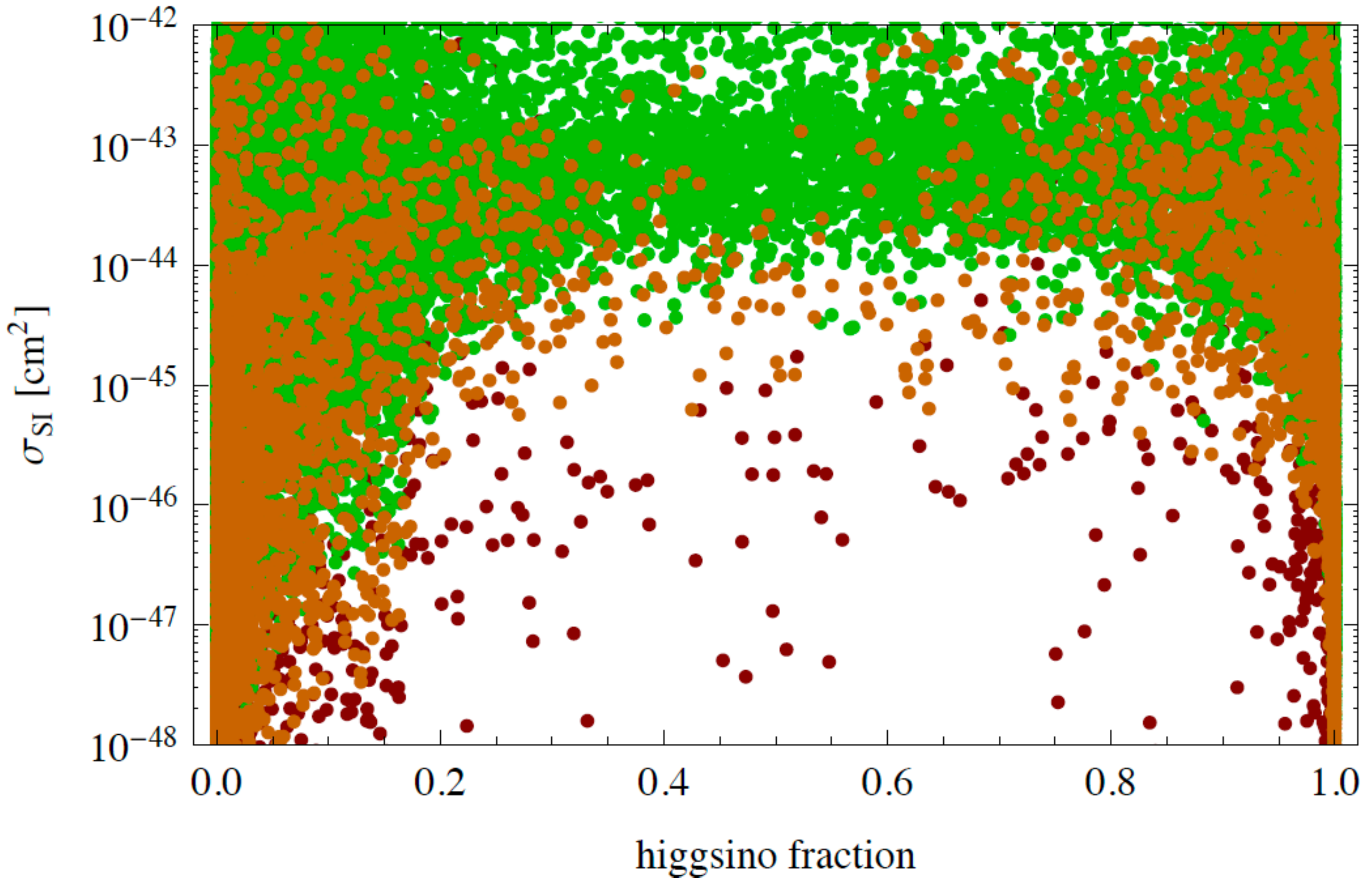}
}
\caption{Direct detection cross section vs. Higgsino fraction of the neutralino. Green, orange and red points correspond to $\Delta_{\rm acc}$ below 10, between 10 and 30, and above 30, respectively.
}
\label{fig:hfractionALL}
\end{center}
\end{figure}

\begin{figure}[th]
\begin{center}
\centerline {
\includegraphics[width=3.3in]{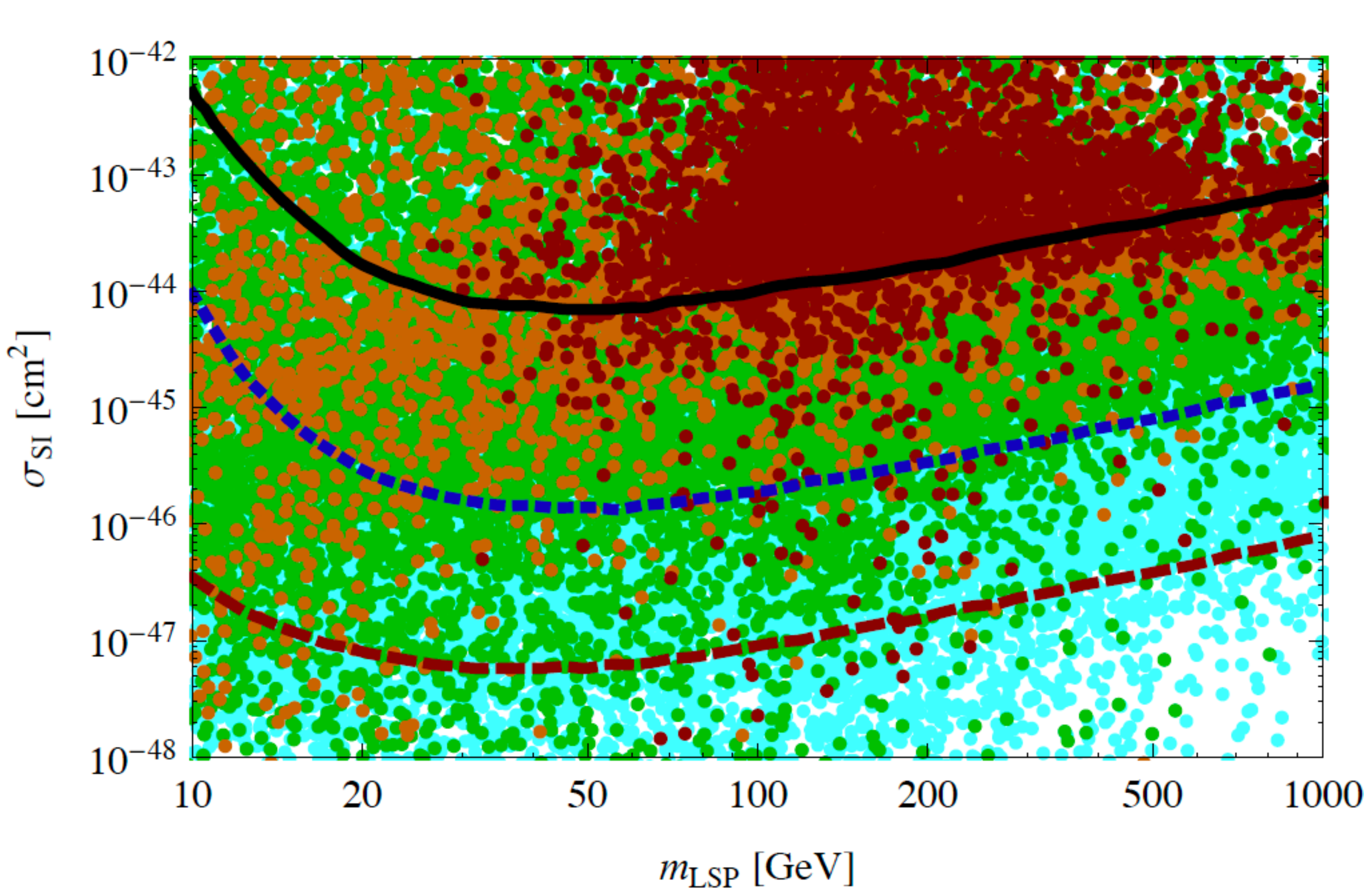}
\includegraphics[width=3.3in]{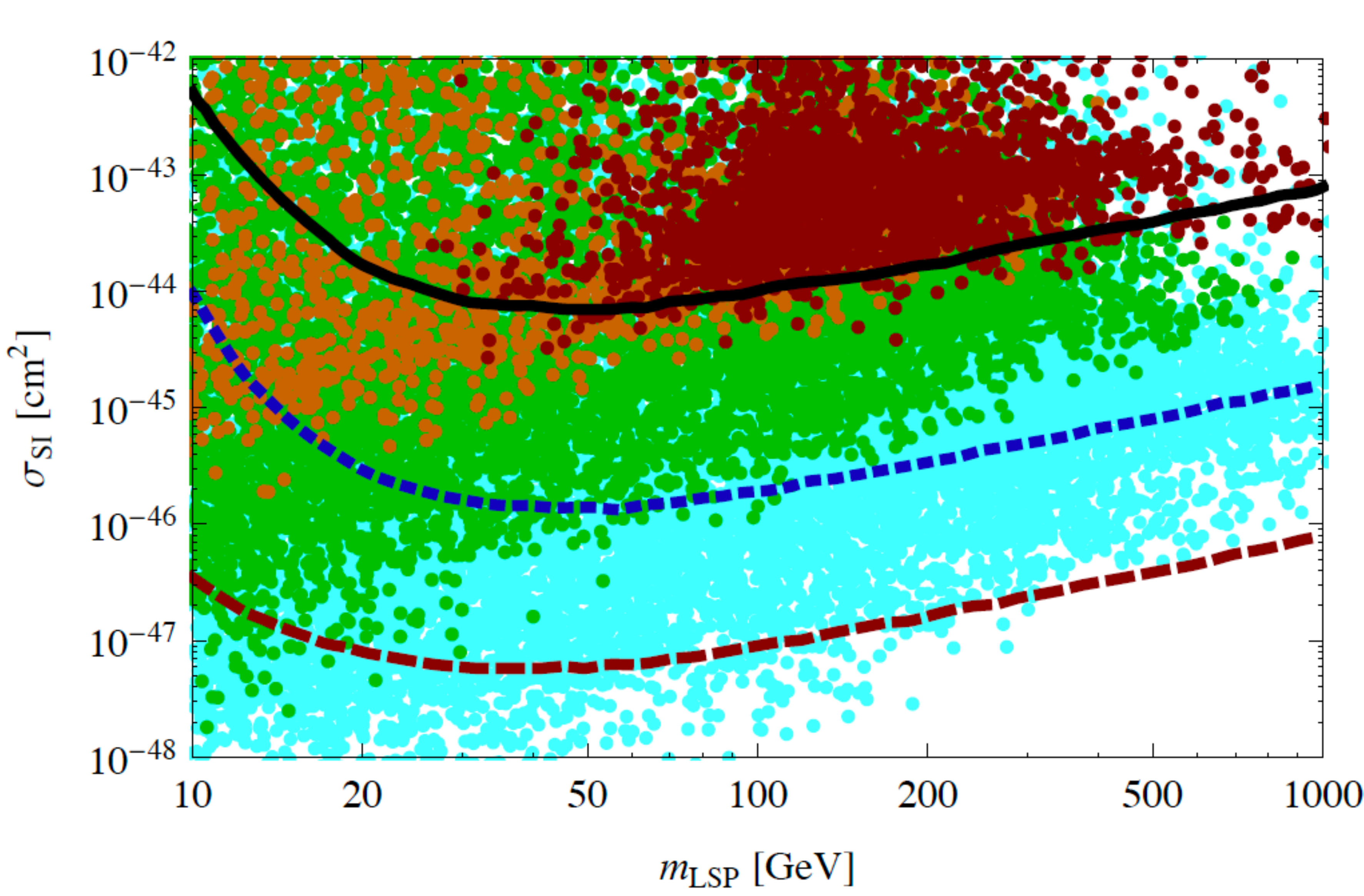}
}
\caption{Left panel: Direct detection cross section vs. the dark matter particle mass, for points with purity above 0.2 (red), between 0.1 and 0.2 (orange), 0.01 and 0.1 (green), and $10^{-3}$ and 0.01 (cyan). Current and projected XENON bounds are superimposed (notation as in Fig.~\ref{fig:hfraction}). Right panel: Same, including only points with $\Delta_{\rm acc}\leq 5 $. }
\label{fig:hfractionMASS}
\end{center}
\end{figure}

Fig.~\ref{fig:hfractionALL} demonstrates the correlation between the direct detection cross section and the Higgsino fraction of the neutralino in this sample. The bulk of points show a correlation similar to that in the real-only sample: Most points with very small direct-detection cross sections are close to either pure-gaugino or pure-Higgsino limit. However, there are some ``outliers", which have direct detection cross sections well below $10^{-44}$ cm$^2$ in spite of having order-one Higgsino and gaugino fractions. The reason for this is accidental cancellations within the $t$-channel contribution to the cross section. An accidental cancellation in the cross section at a particular point in the parameter space is characterized by its anomalous sensitivity to the input Lagrangian parameters at that point. To quantify this, we introduce the measure
\beq
\Delta_{\rm acc} \equiv \sqrt{\sum_{i=1}^5 \left( \frac{\partial \log \sigma}{\partial \log p_i} \right)^2}\,,
\eeq{Delta_acc}   
where $p_i=(M_1, M_2, \mu, \tan\beta, m_A)$. The points in Fig.~\ref{fig:hfractionALL} are color-coded according to this measure, making it clear that the outlier points are uniformly characterized by 
severe accidental cancellations. Thus, the conclusion of the analysis in the previous section remains unchanged: Limits below $10^{-44}$ cm$^2$ imply that the MSSM dark matter neutralino has to be either pure gaugino or pure Higgsino, {\it unless} there are accidental cancellations in the direct detection cross section. This is further illustrated by Fig.~\ref{fig:hfractionMASS}. No firm conclusions on the Higgsino fraction can be drawn from the current or projected XENON100 bounds if {\it all} points in the scan are included (left panel). However, once the points with significant accidental cancellations are excluded (right panel), the situation becomes similar to the case of positive-only parameters (compare with Fig.~\ref{fig:hfraction}), and generic order-one Higgsino-gaugino mixture is already strongly disfavored.

\begin{figure}[t]
\begin{center}
\centerline {
\includegraphics[width=4in]{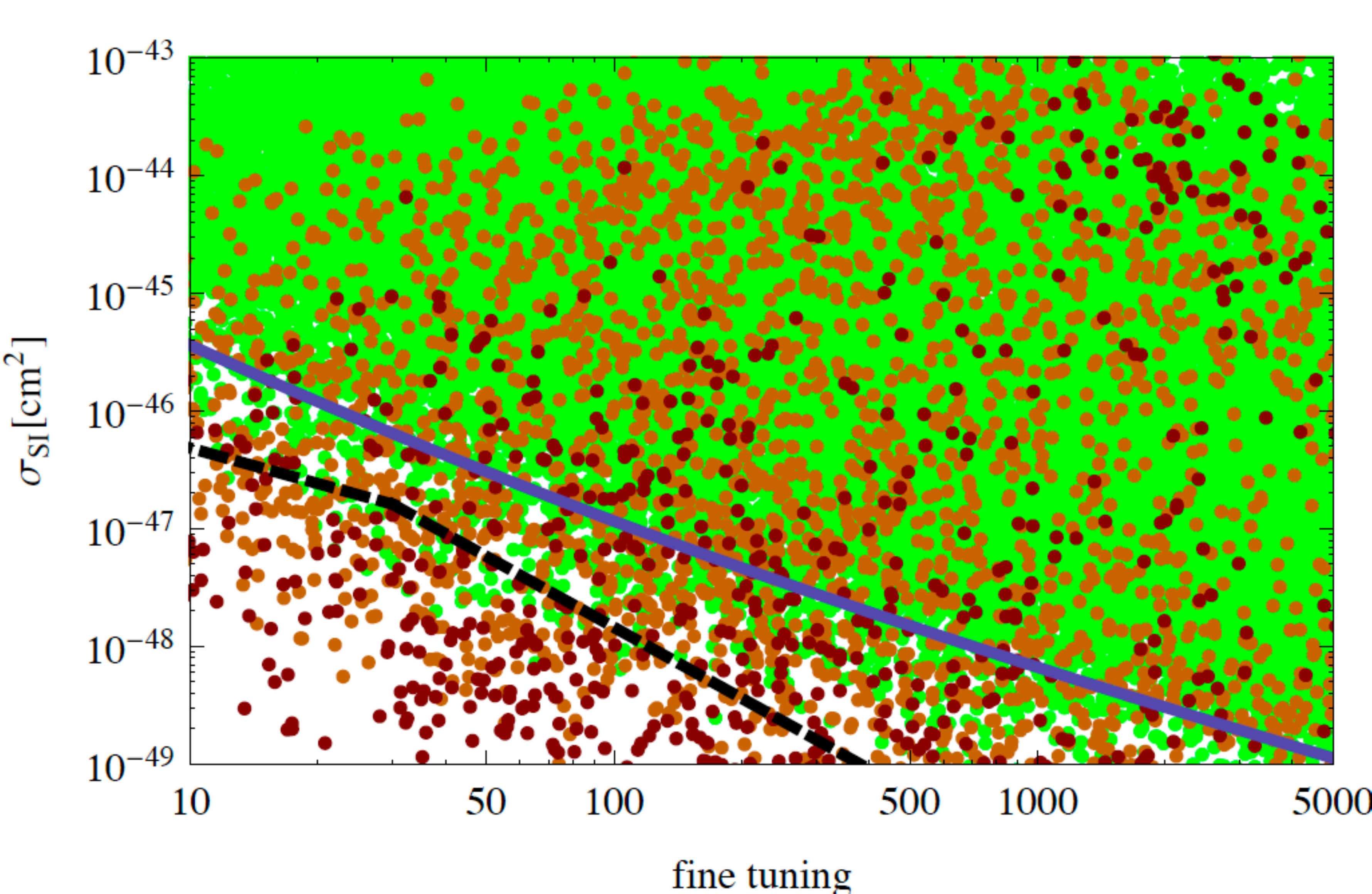}
}
\caption{Direct detection cross section vs. EWSB fine-tuning. Green, orange and red points correspond to $\Delta_{\rm acc}$ below 10, between 10 and 30, and above 30, respectively. Also shown is the analytic lower bound in Eq.~\leqn{FTanal} (blue/solid line), and its modified version in Eq. ~\leqn{FTanal1} (black/dashed line).
}
\label{fig:EWtuningALL}
\end{center}
\end{figure}

\begin{figure}[th]
\begin{center}
\centerline {
\includegraphics[width=3.3in]{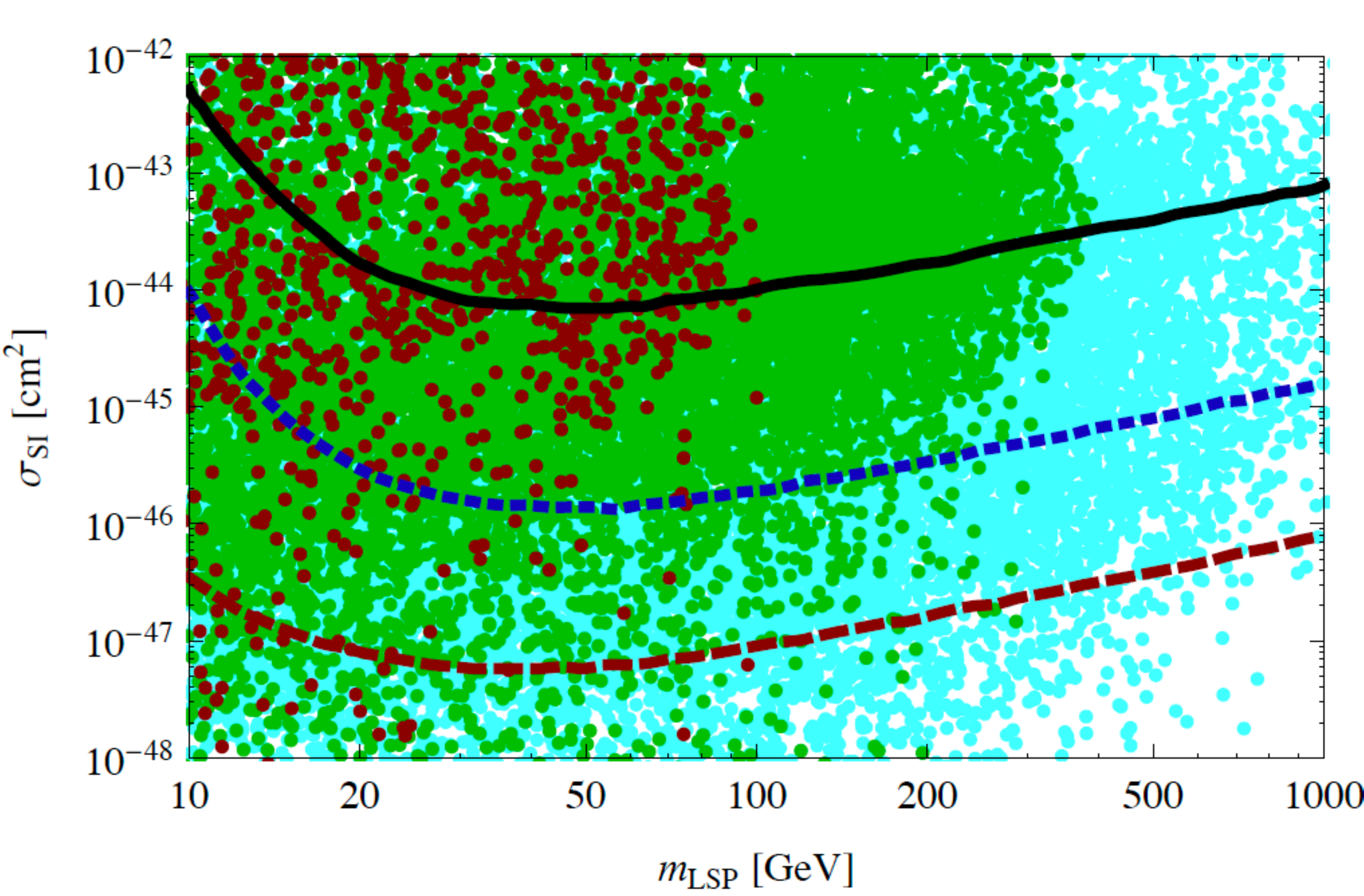}
\includegraphics[width=3.3in]{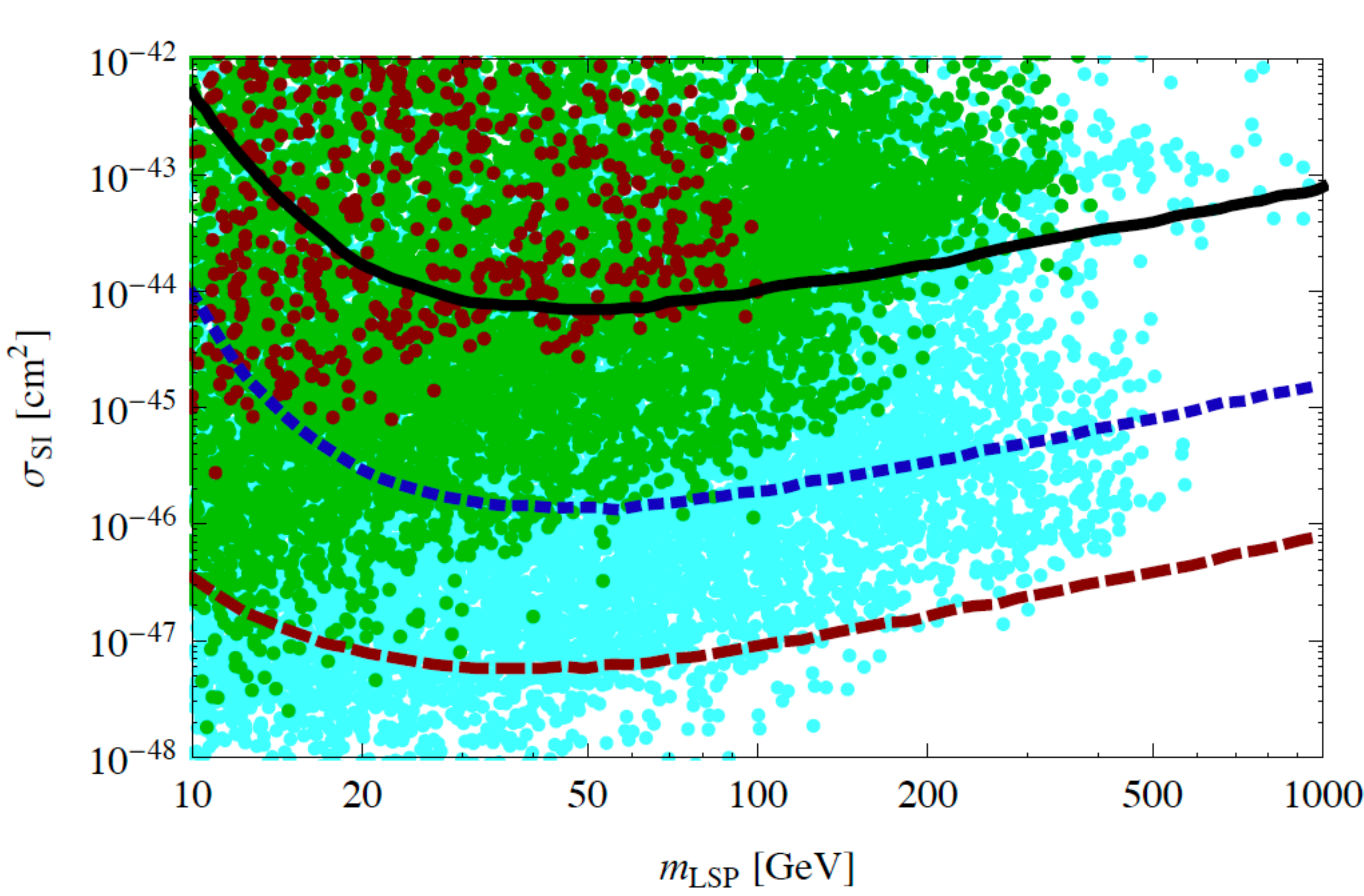}
}
\caption{Left panel: Direct detection cross section vs. dark matter particle mass, for gaugino-like neutralino.
Red, green and cyan points correspond to EWSB fine-tuning in the intervals $(0, 10)$; $[10, 100)$; and$[100,1000]$, respectively. Right panel: Same, including only points with $\Delta_{\rm acc}\leq 5 $. Current and projected XENON bounds are superimposed (notation as in Fig.~\ref{fig:hfraction}).}
\label{fig:MLSP_tuningALL}
\end{center}
\end{figure}

A similar picture emerges for the correlation between direct detection cross section and EWSB fine-tuning in the gaugino LSP case (see Sec.~\ref{sec:EWSB_FT}). Fig.~\ref{fig:EWtuningALL} shows this correlation. The vast majority of points in the scan obey the lower bound on the cross section as a function of $\Delta$, Eq.~\leqn{FTanal}. Most points that violate the bound, and all that violate it by an order of magnitude or more, suffer from severe accidental cancellations. In fact, a slightly stronger version of the analytic lower bound,
\beq
\sigma_{\rm min} \,=\, (1.2\times 10^{-42}~{\rm cm}^2) \,\left( \frac{120~{\rm GeV}}{m_h}\right)^4 \frac{1}{\Delta}\,\left( \min\left[ \frac{1}{\tan\beta}, \frac{1}{\sqrt{\Delta}}\frac{M_{\rm LSP}}{m_Z}\right] \right)^2\,.
\eeq{FTanal1}
is obeyed by all but a few of models with no strong accidental cancellations. 

The direct detection/EWSB fine-tuning correlation is further illustrated by Fig.~\ref{fig:MLSP_tuningALL}. If all points are included (left panel), the correlation is somewhat washed out, with a fairly large number of ``natural" MSSM points yielding cross sections well below current or projected XENON bounds, and even some points with cross sections below the neutrino background of $10^{-48}$ cm$^2$~\cite{nus}. Most of these points, however, are characterized by severe accidental cancellations, and once they are removed (right panel), the correlation between cross section and EWSB tuning is almost as clear as in the positive-only case (compare with Fig.~\ref{fig:MLSP_tuning}). Thus, once points with accidental cancellations are discarded, the discussion of fine-tuning implications of cross section bounds at the end of Sec.~\ref{sec:EWSB_FT} qualitatively applies to the more general case where MSSM soft masses of either sign are allowed, although quantitative statements are slightly weaker: for example, a few points with EWSB fine-tuning of 1 part in 10 or less and LSP masses in the 70--100 GeV range survive the current XENON100 bound. 

\begin{figure}[th]
\begin{center}
\centerline {
\includegraphics[width=3.3in]{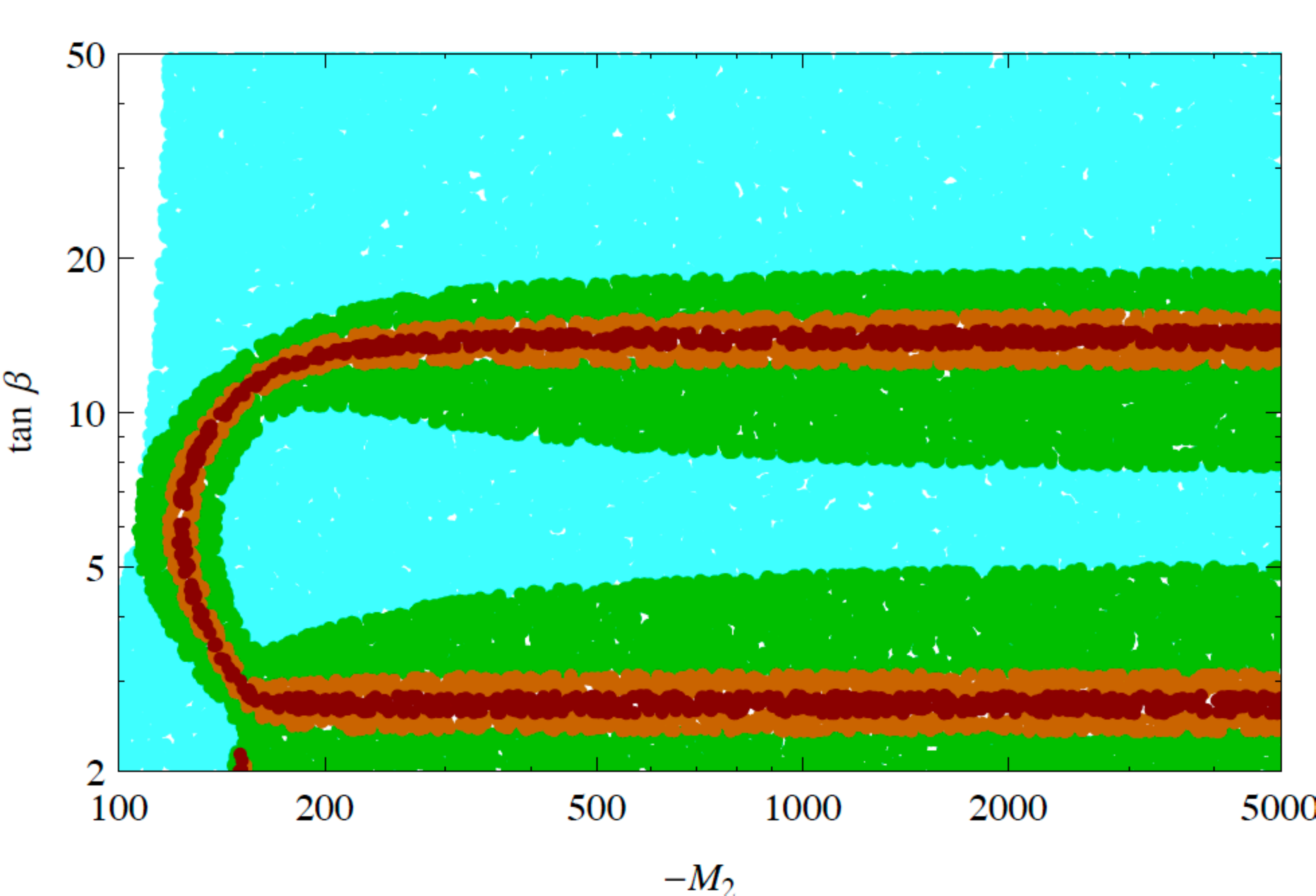}
\includegraphics[width=3.4in]{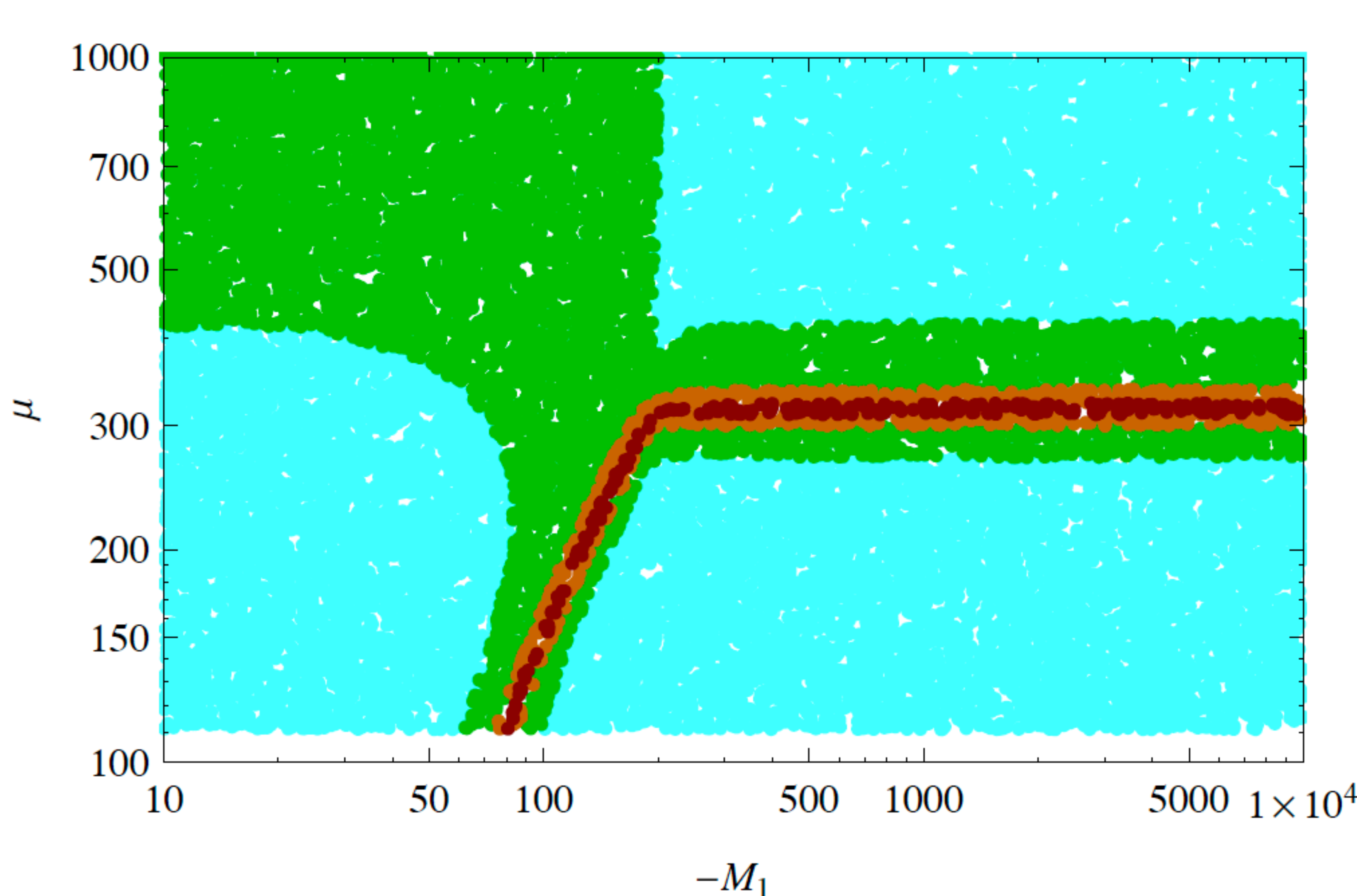}
}
\caption{Left panel: Scatter plot of direct detection cross section as a function of $\tan\beta$ and $-M_2$, with $M_1=-150$ GeV, $\mu=200$ GeV, and $m_A=500$ GeV. Cyan, green, orange and red points have $\log_{10}\sigma_{{\rm cm}^2}>-45, \log_{10}\sigma_{{\rm cm}^2}\in(-46,-45), \log_{10}\sigma_{{\rm cm}^2}\in(-46,-47)$ and $\log_{10}\sigma_{{\rm cm}^2}<-47$, respectively. Right panel: Same, as a function of $\mu$ and $-M_1$, with $M_2=-200$ GeV, $\tan\beta=10$, and $m_A=500$ GeV. }
\label{fig:acc_plots}
\end{center}
\end{figure}

A further demonstration of the accidental nature of cancellations leading to low direct-detection cross sections is given in Fig.~\ref{fig:acc_plots}, where we fix three of the five MSSM parameters and scan over the other two. Points with low cross sections (orange, below $10^{-46}$ cm$^2$, and red, below $10^{-47}$ cm$^2$) are clearly seen to be confined to narrow bands within the parameter space.  

\begin{figure}[t]
\begin{center}
\centerline {
\includegraphics[width=4in]{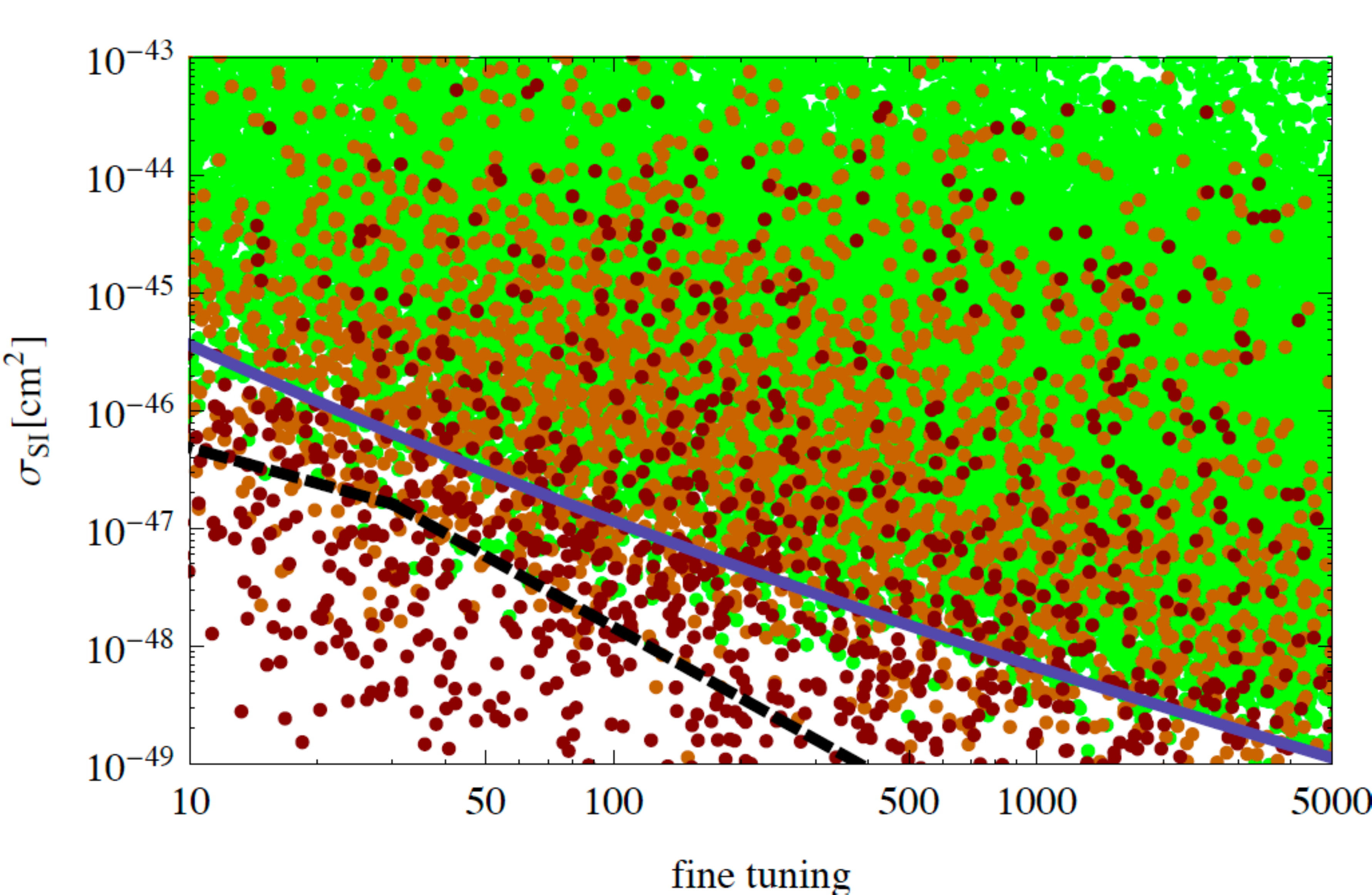}
}
\caption{Direct detection cross section vs. EWSB fine-tuning, for the scan with arbitrary phases of $M_1$ and $M_2$. The notation is identical to Fig.~\ref{fig:EWtuningALL}.
}
\label{fig:complex}
\end{center}
\end{figure}

Finally, we repeated the scan allowing $M_1$ and $M_2$ to have random phases between $0$ and $2\pi$. This may be allowed if squarks and sleptons are very heavy, suppressing contributions to EDMs~\cite{bigphases}. The results of the scan are broadly similar to those for real parameters. Again, the correlations observed in section~\ref{sec:partial} persist, but are somewhat obscured by points with accidental cancellations. As an example, in Fig.~\ref{fig:complex} we show a scatter plot of direct detection cross section vs. EWSB fine-tuning for models with a gaugino-like LSP. We use $\Delta_{\rm acc}$ to quantify accidental cancellations, extending its definition to include the logarithmic derivatives with respect to the two phases in the sum. Again, the vast majority of points obey the lower bound on the cross section as a function of fine-tuning, and the ones that do not typically have $\Delta_{\rm acc}\gg 1$, indicating accidental cancellations.

\section{Conclusions}
\label{sec:conc}

In this paper, we considered the theoretical implications of direct dark matter searches in the MSSM framework. Unlike almost all previous analyses, we did not require any relations among the MSSM parameters, such as mSUGRA, and did not impose thermal relic density constraints. Instead, we make the MSSM parameter space tractable by assuming that no accidental cancellations take place among the $s$-channel and $t$-channel contributions to the spin-independent elastic neutralino-quark cross section, and focusing on the $t$-channel diagrams. These only depend on 5 MSSM parameters. We performed extensive scans over these parameters, in order to identify correlations between the direct detection cross sections and other quantities of physical importance. The following simple picture emerges from our analysis: 

\begin{itemize}

\item If the LSP is a generic mixture of Higgsino and gauginos with order-one admixture of each component, the direct detection cross section is always above $2\times 10^{-44}$ cm$^2$. Such models are already severely constrained: for example, essentially all models with a Higgsino fraction between 0.2 and 0.8 are ruled out by XENON100. 

\item Lowering direct detection cross section below the $10^{-44}$ cm$^2$ level requires that the LSP be either pure gaugino (bino or wino), or pure Higgsino. In both cases, smaller cross sections correlate with higher purity ({\it i.e.}, smaller admixture of the subdominant components) of the LSP.

\item If the LSP is a gaugino, smaller direct detection cross sections correlate with stronger fine-tuning in the electroweak symmetry breaking sector, since they require higher values of the $\mu$ parameter. The current XENON100 bound already implies non-trivial fine-tuning, albeit rather mild at this point: for example, 1 part in 10 or worse tuning is required if the LSP mass is above 70 GeV. Future experiments could put stronger stress on the model: for example, if no signal is seen at the proposed XENON100 upgrade, fine-tuning of 1\% or worse would be required for LSP mass above 100 GeV. This would present a new ``little hierarchy problem" for the MSSM (at least if the idea that the neutralino makes up all of the dark matter is taken seriously). 

\item If the LSP is a Higgsino, no correlation between direct detection cross section and fine-tuning can be established. However, if an additional mild assumption, the one-sided thermal relic density constraint (see sec.~\ref{sec:higgsino}) is imposed, all points with direct detection cross section below about  $2\times 10^{-44}$ cm$^2$ have an LSP mass of 1 TeV or above, and electroweak fine-tuning at the level of $0.25$\% or worse. Non-observation of a signal at the proposed XENON100 upgrade would imply this level of tuning in the Higgsino LSP scenario.

\end{itemize}

Once again, it is worth emphasizing that these conclusions are very general, and apply in the full ``phenomenological" MSSM, without undue theoretical prejudice~\cite{no_prej}. The degree of fine-tuning in the EWSB is a widely accepted quantitative ``figure of merit" used to assess relative theoretical attractiveness of various regions of the MSSM parameter space. We established a clear correlation between this measure and the direct detection cross section. The main conclusion of our analysis is that, if the MSSM is the true model of microscopic physics and dark matter, and no fine-tuning at a sub-percent level is present, the direct dark matter searches currently being conducted and designed should lead to a discovery.

A similar analysis could be performed in other models of electroweak symmetry breaking with particle dark matter candidates. An interesting direction for future work is the so-called next-to-minimal supersymmetric standard model (NMSSM), an MSSM with an additional singlet field in the Higgs sector. This model significantly alleviates the fine-tuning due to the Higgs mass lower bound, and the additional ``singlino" admixture may be present in the LSP, affecting the direct detection cross section. It would be interesting to see if the correlations discussed here persist in the NMSSM. 

\vskip0.8cm
\noindent{\large \bf Acknowledgments} 
\vskip0.3cm
This research is supported by the U.S. National Science Foundation through grant PHY-0757868 and CAREER grant No. PHY-0844667.  This research was  supported in part by the National Science Foundation under Grant No. PHY05-51164. MP is grateful to Michael Peskin and to Aaron Pierce for useful remarks.
MP acknowledges the hospitality of the Kavli Institute for Theoretical Physics at Santa Barbara, where part of this work has been completed. BS acknowledges the hospitality of the Theoretical Advanced Study Institute (TASI-11) at the Univeristy of Colorado at Boulder.

\begin{appendix}

\section{Analytic Direct Detection/Fine-Tuning Relation for Gaugino LSP}
\label{app:FTanal}

In this appendix, we will derive the formula~\leqn{FTanal} for the minimal direct detection cross section consistent with fixed values of $\tan\beta$, $M_{\rm LSP}$, and $\Delta$. We assume real and positive $\mu$, $M_1$ and $M_2$, as in the partial scan of section~\ref{sec:partial}, and focus on the gaugino LSP region, defined by $\mu>M_1$ and/or $\mu>M_2$. As discussed in section~\ref{sec:partial}, a small direct detection cross section requires a {\it pure} gaugino LSP, which implies $\mu\gg M_1$ and/or $\mu\gg M_2$, and $Z_{\chi3}, Z_{\chi4}\ll 1$. Moreover, it is clear from Eq.~\leqn{nnh} that the direct detection cross section is minimized if the LSP is predominantly bino, since $g^\prime<g$. Motivated by these considerations, consider the limit 
\beq
M_1 \ll \mu \ll M_2\,.
\eeq{limit}
Ignoring terms suppressed by $M_2$, the neutralino mixing angles are obtained by diagonalizing the mass matrix
\beq
M_{\tilde{\chi}^0} \,=\, \left(
\begin{array}{ccc}
M_1 & -c_\beta s_w m_Z & s_\beta s_w m_Z\\
-c_\beta s_w m_Z & 0 & -\mu\\
s_\beta s_w m_Z & -\mu & 0 
\end{array}\right)\,,
\eeq{mass_mat}
where $s_\theta\equiv\sin\theta$, $c_\theta\equiv\cos\theta$, and subscript $w$ denotes the Weinberg angle. Assuming $m_Z\ll \mu$ (which is always the case in the region of interest) and working to second order in $1/\mu$, we obtain
\beqa
Z_{\chi3} &\approx& s_w \frac{m_Z}{\mu}\,\left( s_\beta + c_\beta \frac{M_1}{\mu} \right)\,,\CR
Z_{\chi4} &\approx& s_w \frac{m_Z}{\mu}\,\left( -c_\beta - s_\beta \frac{M_1}{\mu} \right)\,.
\eeqa{approxZs}
Barring accidental cancellations, the direct-detection cross section is minimized in the Higgs decoupling limit, $m_A\gg m_Z$. In this limit, the well-known relationships
\beq
\frac{\sin2\alpha}{\sin2\beta} = - \frac{m_H^2+m_h^2}{m_H^2-m_h^2} \rightarrow -1\,,~~~\frac{\tan2\alpha}{\tan2\beta} = \frac{m_A^2+m_Z^2}{m_A^2-m_Z^2}\rightarrow 1
\eeq{alpha_beta}
imply $\sin\alpha\approx-\cos\beta$, $\cos\alpha\approx\sin\beta$. Plugging this and Eq.~\leqn{approxZs} into Eqs.~\leqn{A_i},~\leqn{defs_up}, and~\leqn{defs_d}, and ignoring the heavy Higgs exchange diagrams, yields\footnote{The approximate expression for the direct detection cross section in the bino-LSP limit also appeared in Ref.~\cite{KN}.}
\beq
A_u \,\approx\, A_d \,\approx\, \frac{\pi \alpha}{c_w^2}\frac{1}{m_h^2\mu}\,\left( \sin 2\beta + \frac{M_{\rm LSP}}{\mu} \right)\,,
\eeq{As_mu}
where $\alpha$ is the fine structure constant, and we used $M_{\rm LSP} = M_1 + O(1/\mu)$. The final step is to relate $\mu$ to the fine-tuning measure $\Delta$. For $\mu>0$, the two terms in the brackets in Eq.~\leqn{As_mu} cannot cancel, and the direct detection cross section is minimized (for fixed $\mu$ and $M_{\rm LSP}$) at large $\tan\beta$, since $\sin2\beta \sim 2\tan^{-1}\beta \to 0$ in this limit. As discussed in section~\ref{sec:setup}, the fine-tuning in this limit is dominated by $\delta(\mu)$ and $\delta(b)$, see Eq.~\leqn{conlarget}, so that
\beq
\Delta\approx \frac{4}{m_Z^2}\sqrt{\mu^4+\frac{m_A^4}{\tan^2\beta}}\,.
\eeq{Dlargetanb}
It is clear that the smallest direct detection cross section for a given value of $\Delta$ is achieved for the largest possible $\mu$ consistent with that $\Delta$, which occurs when $m_A \ll \sqrt{\tan\beta} \mu$. 
(It can be easily checked that there are always values of $m_A$ consistent with this condition which are still large enough to ignore the $H$ exchange contributions to direct detection matrix elements.) In this situation,
\beq
\Delta \approx \frac{4\mu^2}{m_Z^2}\,,
\eeq{Dopt}
so that we can rewrite Eq.~\leqn{As_mu} in its final form 
\beq
A_u \,\approx\, A_d \,\approx\, \frac{4\pi \alpha}{c_w^2 m_Z m_h^2} \,\frac{1}{\sqrt{\Delta}}\,\left( \frac{1}{\tan\beta} \,+\, \frac{1}{\sqrt{\Delta}}\frac{M_{\rm LSP}}{m_Z}\right).
\eeq{Aanal}
Squaring and multiplying by the appropriate nuclear formfactors yields Eq.~\leqn{FTanal}.

\end{appendix}


\begin{thebibliography}{99}

\bibitem{SUSY_DM}
  H.~Goldberg,
  Phys.\ Rev.\ Lett.\  {\bf 50}, 1419 (1983)
  [Erratum-ibid.\  {\bf 103}, 099905 (2009)];\\
 J.~R.~Ellis, J.~S.~Hagelin, D.~V.~Nanopoulos, K.~A.~Olive and M.~Srednicki,
  Nucl.\ Phys.\  B {\bf 238}, 453 (1984);\\
  M.~Drees and M.~M.~Nojiri,
  Phys.\ Rev.\  D {\bf 47}, 376 (1993)
  [arXiv:hep-ph/9207234].
  
\bibitem{DMreviews}
For reviews, see for example
  G.~Jungman, M.~Kamionkowski and K.~Griest,
  Phys.\ Rept.\  {\bf 267}, 195 (1996)
  [arXiv:hep-ph/9506380];\\
  G.~Bertone, D.~Hooper and J.~Silk,
  Phys.\ Rept.\  {\bf 405}, 279 (2005)
  [arXiv:hep-ph/0404175];\\
 L.~Bergstrom,
  New J.\ Phys.\  {\bf 11}, 105006 (2009)
  [arXiv:0903.4849 [hep-ph]];\\
 J.~L.~Feng,
  Ann.\ Rev.\ Astron.\ Astrophys.\  {\bf 48}, 495 (2010)
  [arXiv:1003.0904 [astro-ph.CO]].
  
\bibitem{cdms}
  Z.~Ahmed {\it et al.}  [The CDMS-II Collaboration],
  Science {\bf 327}, 1619 (2010)
  [arXiv:0912.3592 [astro-ph.CO]].

\bibitem{edel}
  E.~Armengaud {\it et al.}  [EDELWEISS Collaboration],
 {\it ``Final results of the EDELWEISS-II WIMP search using a 4-kg array of
  cryogenic germanium detectors with interleaved electrodes,''}
  arXiv:1103.4070 [astro-ph.CO].

\bibitem{x100old}
  E.~Aprile {\it et al.}  [XENON100 Collaboration],
  Phys.\ Rev.\ Lett.\  {\bf 105}, 131302 (2010)
  [arXiv:1005.0380 [astro-ph.CO]].
  
\bibitem{x100}
  E.~Aprile {\it et al.}  [XENON100 Collaboration],
{\it ``Dark Matter Results from 100 Live Days of XENON100 Data,''}
  arXiv:1104.2549 [astro-ph.CO].

\bibitem{dama}
  R.~Bernabei {\it et al.}  [DAMA Collaboration],
  Eur.\ Phys.\ J.\  C {\bf 56}, 333 (2008)
  [arXiv:0804.2741 [astro-ph]].

\bibitem{cogent}
  C.~E.~Aalseth {\it et al.}  [CoGeNT collaboration],
  Phys.\ Rev.\ Lett.\  {\bf 106}, 131301 (2011)
  [arXiv:1002.4703 [astro-ph.CO]].

\bibitem{GPMM}
  V.~Mandic, A.~Pierce, P.~Gondolo and H.~Murayama,
{\it ``The Lower bound on the neutralino nucleon cross-section,''}
  arXiv:hep-ph/0008022.

\bibitem{sa}
  R.~Kitano, Y.~Nomura,
  Phys.\ Lett.\  {\bf B632}, 162-166 (2006)
  [hep-ph/0509221];\\
 T.~Cohen, D.~J.~Phalen, A.~Pierce,
  Phys.\ Rev.\  {\bf D81}, 116001 (2010)
  [arXiv:1001.3408 [hep-ph]].
  
\bibitem{DG}
  S.~Cassel, D.~M.~Ghilencea and G.~G.~Ross,
  Phys.\ Lett.\ B {\bf 687}, 214 (2010)
  [arXiv:0911.1134 [hep-ph]];\\
  S.~Cassel, D.~M.~Ghilencea and G.~G.~Ross,
  Nucl.\ Phys.\ B {\bf 835}, 110 (2010)
  [arXiv:1001.3884 [hep-ph]];\\
  S.~Cassel, D.~M.~Ghilencea, S.~Kraml, A.~Lessa and G.~G.~Ross,
  JHEP {\bf 1105}, 120 (2011)
  [arXiv:1101.4664 [hep-ph]].
    
\bibitem{KN}
  R.~Kitano and Y.~Nomura,
{\it ``Supersymmetry with Small mu: Connections between Naturalness, Dark Matter,
  and (Possibly) Flavor,''}
  arXiv:hep-ph/0606134.
  
 \bibitem{FS}
  J.~L.~Feng and D.~Sanford,
  JCAP {\bf 1105}, 018 (2011)
  [arXiv:1009.3934 [hep-ph]].
  
  \bibitem{dmdensity}
  M.~Pato, O.~Agertz, G.~Bertone, B.~Moore, R.~Teyssier,
  Phys.\ Rev.\  {\bf D82}, 023531 (2010).
  [arXiv:1006.1322 [astro-ph.HE]];\\
  M.~Kamionkowski, S.~M.~Koushiappas,
  Phys.\ Rev.\  {\bf D77}, 103509 (2008).
  [arXiv:0801.3269 [astro-ph]].
  
  \bibitem{veldist}
  C.~McCabe,
  Phys.\ Rev.\  {\bf D82}, 023530 (2010).
  [arXiv:1005.0579 [hep-ph]];  \\
  A.~M.~Green,
  JCAP {\bf 1010}, 034 (2010).
  [arXiv:1009.0916 [astro-ph.CO]].

 
\bibitem{isospin}
  F.~Giuliani,
  Phys.\ Rev.\ Lett.\  {\bf 95}, 101301 (2005).
  [hep-ph/0504157];\\
  J.~L.~Feng, J.~Kumar, D.~Marfatia, D.~Sanford,
  Phys.\ Lett.\  {\bf B703}, 124-127 (2011).
  [arXiv:1102.4331 [hep-ph]].
  
\bibitem{akula} 
  S.~Akula, D.~Feldman, Z.~Liu, P.~Nath and G.~Peim,
  Mod.\ Phys.\ Lett.\ A {\bf 26}, 1521 (2011)
  [arXiv:1103.5061 [hep-ph]].
  
\bibitem{strumia}
  M.~Farina, M.~Kadastik, D.~Pappadopulo, J.~Pata, M.~Raidal and A.~Strumia,
{\it ``Implications of XENON100 results for Dark Matter models and for the LHC,''}
  arXiv:1104.3572 [hep-ph].

\bibitem{xenon_msugra}
  S.~Profumo,
{\it ``The Quest for Supersymmetry: Early LHC Results versus Direct and Indirect
  Neutralino Dark Matter Searches,''}
  arXiv:1105.5162 [hep-ph];\\
  O.~Buchmueller {\it et al.},
 {\it ``Supersymmetry and Dark Matter in Light of LHC 2010 and Xenon100 Data,''}
  arXiv:1106.2529 [hep-ph];\\
  G.~Bertone, D.~G.~Cerdeno, M.~Fornasa, R.~R.~de Austri, C.~Strege and R.~Trotta,
 {\it ``Global fits of the cMSSM including the first LHC and XENON100 data,''}
  arXiv:1107.1715 [hep-ph].
 
\bibitem{nucs}
  J.~R.~Ellis, K.~A.~Olive and C.~Savage,
  Phys.\ Rev.\  D {\bf 77}, 065026 (2008)
  [arXiv:0801.3656 [hep-ph]].

\bibitem{As}
  T.~Falk, A.~Ferstl and K.~A.~Olive,
  Phys.\ Rev.\  D {\bf 59}, 055009 (1999)
  [Erratum-ibid.\  D {\bf 60}, 119904 (1999)]
  [arXiv:hep-ph/9806413].

\bibitem{Martin}
  S.~P.~Martin,
{\it ``A supersymmetry primer,''}
  arXiv:hep-ph/9709356.

\bibitem{EDMs}
  M.~Pospelov and A.~Ritz,
  Annals Phys.\  {\bf 318}, 119 (2005)
  [arXiv:hep-ph/0504231].

\bibitem{bigphases}
  N.~Arkani-Hamed, S.~Dimopoulos, G.~F.~Giudice and A.~Romanino,
  Nucl.\ Phys.\  B {\bf 709}, 3 (2005)
  [arXiv:hep-ph/0409232];\\
  N.~G.~Deshpande and J.~Jiang,
  Phys.\ Lett.\  B {\bf 615}, 111 (2005)
  [arXiv:hep-ph/0503116];\\
 D.~Chang, W.~F.~Chang and W.~Y.~Keung,
  Phys.\ Rev.\  D {\bf 71}, 076006 (2005)
  [arXiv:hep-ph/0503055];\\
  G.~F.~Giudice and A.~Romanino,
  Phys.\ Lett.\  B {\bf 634}, 307 (2006)
  [arXiv:hep-ph/0510197].

\bibitem{golden}
  M.~Perelstein and C.~Spethmann,
  JHEP {\bf 0704}, 070 (2007)
  [arXiv:hep-ph/0702038].

\bibitem{BG}
  R.~Barbieri and G.~F.~Giudice,
  Nucl.\ Phys.\ B {\bf 306}, 63 (1988).

\bibitem{LH}
  R.~Barbieri and A.~Strumia,
{\it ``The 'LEP paradox',''}
  arXiv:hep-ph/0007265.

\bibitem{x100F}
E. Aprile [Xenon Collaboration], J. Phys. Conf. Ser. 203 (2010) 012005.

\bibitem{xtalk}
{\tt http://www-conf.slac.stanford.edu/tevpa09/Santorelli090713v2.pdf}

\bibitem{xmass}
  H.~Sekiya [XMASS~collaboration],
{\it ``Xmass,''} [arXiv:1006.1473 [astro-ph.IM]].

\bibitem{lux}
  S.~Fiorucci {\it et al.},
  AIP Conf.\ Proc.\  {\bf 1200}, 977-980 (2010).
  [arXiv:0912.0482 [astro-ph.CO]].

\bibitem{supercdms}
 T.~Bruch [CDMS~Collaboration],
{\it ``CDMS-II to SuperCDMS: WIMP search at a zeptobarn,''}
  [arXiv:1001.3037 [astro-ph.IM]].

\bibitem{welltemp}
  N.~Arkani-Hamed, A.~Delgado and G.~F.~Giudice,
  Nucl.\ Phys.\  B {\bf 741}, 108 (2006)
  [arXiv:hep-ph/0601041].

\bibitem{nus}
  L.~E.~Strigari,
  New J.\ Phys.\  {\bf 11}, 105011 (2009).
  [arXiv:0903.3630 [astro-ph.CO]];\\
  A.~Gutlein {\it et al.},
  Astropart.\ Phys.\  {\bf 34}, 90-96 (2010).
  [arXiv:1003.5530 [hep-ph]].

\bibitem{PDG}
  K.~Nakamura {\it et al.} [ Particle Data Group Collaboration],
  J.\ Phys.\ G {\bf G37}, 075021 (2010).

\bibitem{higgsinodm}
   K.~Griest, D.~Seckel,
  Phys.\ Rev.\  {\bf D43}, 3191-3203 (1991);\\
   S.~Mizuta, M.~Yamaguchi,
  Phys.\ Lett.\  {\bf B298}, 120-126 (1993).
  [hep-ph/9208251];   \\
    M.~Drees, M.~M.~Nojiri,
  Phys.\ Rev.\  {\bf D47}, 376-408 (1993).
  [hep-ph/9207234];\\
    M.~Drees, M.~M.~Nojiri, D.~P.~Roy, Y.~Yamada,
  Phys.\ Rev.\  {\bf D56}, 276-290 (1997).
  [hep-ph/9701219];\\
  J.~Edsjo, P.~Gondolo,
  Phys.\ Rev.\  {\bf D56}, 1879-1894 (1997).
  [hep-ph/9704361].

\bibitem{darksusy}
  P.~Gondolo, J.~Edsjo, P.~Ullio, L.~Bergstrom, M.~Schelke, E.~A.~Baltz,
  JCAP {\bf 0407}, 008 (2004)
  [astro-ph/0406204]; \\
  P. Gondolo, J. Edsjš, P. Ullio, L. Bergstršm, M. Schelke, E.A. Baltz, T. Bringmann and G. Duda, {\tt http://www.darksusy.org}

\bibitem{no_prej}
  C.~F.~Berger, J.~S.~Gainer, J.~L.~Hewett and T.~G.~Rizzo,
  JHEP {\bf 0902}, 023 (2009)
  [arXiv:0812.0980 [hep-ph]];\\
    J.~S.~Gainer,
  AIP Conf.\ Proc.\  {\bf 1200}, 1015 (2010)
  [arXiv:0910.1375 [hep-ph]].


\end{thebibliography}
\end{document}